\documentclass[oldversion]{aa}  

\usepackage{graphicx}
\usepackage{txfonts}
\usepackage{natbib}
\usepackage{supertabular}
\usepackage{longtable}
\usepackage{url}
\usepackage{dcolumn}
\usepackage{placeins}
\usepackage{multirow}
\usepackage{bm}

\begin{document}
\defcitealias{Prochaska09s}{PW09} 	
\newcommand{\zabs}{\ensuremath{z_{\rm abs}}}
\newcommand{\zem}{\ensuremath{z_{\rm em}}}
\newcommand{\HH}{\mbox{H$_2$}}
\newcommand{\HD}{\mbox{HD}}
\newcommand{\DD}{\mbox{D$_2$}}
\newcommand{\CO}{\mbox{CO}}
\newcommand{\dla}{damped Lyman-$\alpha$}
\newcommand{\Dla}{Damped Lyman-$\alpha$}
\newcommand{\lya}{Ly-$\alpha$}
\newcommand{\lyb}{Ly-$\beta$}
\newcommand{\lyg}{Ly-$\gamma$}
\newcommand{\s}{\ensuremath{\mathcal{S}}}  

\newcommand{\ArI}{\ion{Ar}{i}}
\newcommand{\CI}{\ion{C}{i}}
\newcommand{\CII}{\ion{C}{ii}}
\newcommand{\CIV}{\ion{C}{iv}}
\newcommand{\ClI}{\ion{Cl}{i}}
\newcommand{\ClII}{\ion{Cl}{ii}}
\newcommand{\CrII}{\ion{Cr}{ii}}
\newcommand{\CuII}{\ion{Cu}{ii}}
\newcommand{\DI}{\ion{D}{i}}
\newcommand{\FeII}{\ion{Fe}{ii}}
\newcommand{\HI}{\ion{H}{i}}
\newcommand{\MgI}{\ion{Mg}{i}}
\newcommand{\MgII}{\ion{Mg}{ii}}
\newcommand{\MnII}{\ion{Mn}{ii}}
\newcommand{\NI}{\ion{N}{i}}
\newcommand{\NV}{\ion{N}{v}}
\newcommand{\NiII}{\ion{Ni}{ii}}
\newcommand{\OI}{\ion{O}{i}}
\newcommand{\OVI}{\ion{O}{vi}}
\newcommand{\PII}{\ion{P}{ii}}
\newcommand{\PbII}{\ion{Pb}{ii}}
\newcommand{\SI}{\ion{S}{i}}
\newcommand{\SII}{\ion{S}{ii}}
\newcommand{\SiII}{\ion{Si}{ii}}
\newcommand{\SiIV}{\ion{Si}{iv}}
\newcommand{\TiII}{\ion{Ti}{ii}}
\newcommand{\ZnII}{\ion{Zn}{ii}}
\newcommand{\AlII}{\ion{Al}{ii}}
\newcommand{\AlIII}{\ion{Al}{iii}}

\newcommand{\Ho}{\mbox{$H_0$}}
\newcommand{\angstrom}{\mbox{{\rm \AA}}}
\newcommand{\abs}[1]{\left| #1 \right|} 
\newcommand{\avg}[1]{\left< #1 \right>} 
\newcommand{\fhix}{\ensuremath{f_{\HI}(N,X)}}
\newcommand{\ldla}{\ensuremath{l_{\rm DLA}}}
\newcommand{\omegagdla}{\ensuremath{\Omega_{\rm g}^{\rm DLA}}}
\newcommand{\gz}{\ensuremath{g(z)}}
\newcommand{\zmin}{\ensuremath{z_{\rm min}}}
\newcommand{\zmino}{\ensuremath{z_{\rm min}^0}}
\newcommand{\zmax}{\ensuremath{z_{\rm max}}}
\newcommand{\kms}{\ensuremath{{\rm km\,s^{-1}}}}
\newcommand{\cmsq}{\ensuremath{{\rm cm}^{-2}}}

\title{Evolution of the cosmological mass density of neutral gas from Sloan Digital Sky Survey II - Data Release 7}


   \author{P. Noterdaeme
          \inst{1,2}
          \and
          P. Petitjean 
          \inst{2}
          \and
          C. Ledoux
          \inst{3}
          \and
          R. Srianand
          \inst{1}
          }

\authorrunning{P. Noterdaeme et al.}
\titlerunning{Evolution of the cosmological neutral gas mass density}

   \offprints{P. Noterdaeme}

   \institute{Inter University Centre for Astronomy and Astrophysics, Post Bag 4, Ganesh Khind, Pune 411 007, India\\
              \email{[pasquiern, anand]@iucaa.ernet.in}
\and
UPMC, Universit\'e Paris 6, Institut d'Astrophysique de Paris, CNRS UMR 7095, 98bis bd Arago, 75014 Paris, France\\
              \email{[noterdaeme, petitjean]@iap.fr}
              \and
              European Southern Observatory, Alonso de C\'ordova 3107, Casilla 19001, Vitacura, Santiago 19, Chile\\
              \email{cledoux@eso.org}
              }

   \date{}

  \abstract{
We present the results of a search for damped Lyman-$\alpha$ (DLA) systems in the Sloan Digital Sky Survey II
(SDSS), Data Release 7. We use a fully automatic procedure to identify DLAs and derive their column densities.
The procedure is checked against the results of previous searches for DLAs in SDSS. 
We discuss the agreements and differences and show the robustness of our procedure.
For each system, we obtain an accurate measurement of the absorber's redshift, 
the \HI\ column density and the equivalent width of associated metal absorption lines, without any human 
intervention. We find 1426 absorbers with $2.15<z<5.2$ with $\log N(\HI)\ge20$, out of which 937 
systems have $\log N(\HI)\ge 20.3$. 
This is the largest DLA sample ever built, made available to the scientific community through 
the electronic version of this paper. 

In the course of the survey, we discovered the intervening DLA with highest \HI\ column density known to date
with $\log N(\HI)=22.0\pm0.1$. This single system provides a strong constraint on the high-end of the $N(\HI)$ 
frequency distribution now measured with high accuracy. 

We show that the presence of a DLA at the blue end of a QSO spectrum can lead to important 
systematic errors and propose a method to avoid them.
This has important consequences for the measurement of the cosmological mass density of neutral gas 
at $z\sim2.2$ and therefore on our understanding of galaxy evolution over the past 10 billion 
years.

We find a significant decrease of the cosmological mass density of neutral gas in DLAs, $\omegagdla$, 
from $z=4$ to $z=2.2$, consistent with the result of previous SDSS studies. However, and contrary 
to other SDSS studies, we find that $\omegagdla(z=2.2)$ is about twice the value at $z=0$. 
This implies that $\omegagdla$ keeps decreasing at $z<2.2$.

}
   \maketitle


\section{Introduction}

Despite accounting for only a small fraction of all the baryons in the Universe \citep[see][]{Fukugita04}, 
the neutral and molecular phases of the interstellar medium 
are at any redshift the reservoir of gas from which stars form.
Therefore, determining the cosmological mass density of neutral gas ($\Omega_{\rm HI}$) and
its evolution in time is a fundamental step forward
towards understanding how galaxies form \citep{Klypin95}. 

In the local Universe, neutral gas is best traced by the hyperfine 21-cm emission of atomic 
hydrogen. Its observation allows for an accurate measurement of the neutral gas
spatial distribution in nearby galaxies and strongly constrains the column density frequency distribution 
\citep{Zwaan05} and $\Omega_{\HI}(z=0)$ \citep{Zwaan05b}.
However, the limited sensitivity of current radio telescopes prevents 
direct detections of \HI\ emission beyond $z\sim0.2$ \citep{Lah07,Verheijen07,Catinella08}.

At high redshift, 
most of the neutral hydrogen is revealed by the Damped Lyman-$\alpha$ (DLA) absorption systems 
detected in the spectra of background quasars. While most of the 
gas is likely to be neutral for $\log N(\HI)>19.5$ \citep{Viegas95}, the conventional  
definition for Damped Lyman-$\alpha$ systems is $\log N(\HI)\ge20.3$ \citep{Wolfe86}. Not only 
does this correspond to a convenient detectability limit in low-resolution spectra, but also  
to a critical surface-density limit for star formation.
Since DLAs are easy to detect and the \HI\ column densities can be measured accurately, 
it is possible to measure the cosmological mass density of neutral gas at different redshifts, 
independently of the exact nature of the absorbers, provided a sufficiently large number of 
background quasars is observed. However, any bias affecting the selection of the quasars or the 
determination of the redshift path-length probed by each line of sight can affect the measurements. 

The Lick survey, the first systematic search for DLAs, led to the detection of 15 systems 
at $\avg{z}=2.5$ along the line of sight to 68 quasars \citep{Wolfe86, Turnshek89,Wolfe93}. 
About one hundred quasars were subsequently surveyed for DLA absorptions by \citet{Lanzetta91}. 
A number of surveys have followed \citep[e.g.][]{Wolfe95,Lanzetta95, Storrie-Lombardi96a, Storrie-Lombardi96b, 
Storrie-Lombardi00, Ellison01,Peroux03,Rao05,Rao06}, each of them contributing significantly to increase 
the number of known DLAs and the redshift coverage. 
On the other hand, surveys at low and intermediate redshifts are difficult because they require UV 
observations and the number of confirmed systems is building up very slowly \citep{Lanzetta95,Rao05,Rao06}.

In turn, \citet{Peroux01,Peroux03} aimed at the highest redshifts 
by observing 66 quasars with $\zem>4$. They included data from previous surveys in their analysis, which 
led to the largest DLA sample available at that time. They found no significant evolution of the cosmological 
mass density of neutral gas over the redshift range $1<z<4$ and suggested that a significant part of this mass 
is due to systems with neutral hydrogen column densities below the conventional DLA cutoff limit. 

The most recent contribution to the actual census of DLAs 
is a semi-automatic data mining of thousands of quasar spectra from the Sloan Digital Sky Survey 
\citep{Prochaska04,Prochaska05,Prochaska09s} revealing more than 700 new DLAs thus  
increasing by one order of magnitude the number of known DLAs at $z>2.2$. Key results include the indication that the 
$N(\HI)$-frequency distribution deviates significantly from a single power-law. Its shape is found 
to be nearly invariant with redshift, while its normalisation does change with redshift.
In their work, \citeauthor{Prochaska09s} found that the incidence of DLAs and the neutral gas mass 
density in DLAs ($\omegagdla$) decrease significantly with time between $z\sim3.5$ and $z=2.2$. 
The mass density in DLAs 
at $z\sim2.2$ is claimed to be consistent with that measured at $z=0$ by \citet{Zwaan05}.
However, this is difficult to reconcile with the result from \citet{Rao06} that $\omegagdla$ stays
constant over the range $1 < z < 2$ and then decreases strongly to reach the value derived from 21~cm observations 
at $z=0$. 

\citeauthor{Rao06}'s results using HST could be questioned because DLAs are not searched 
directly but rather are selected on the basis of strong associated \MgII\ absorption. There is indeed an
excess of high column densities in the HST sample compared to the SDSS sample \citep{Rao06}
that could be real or induced by some selection bias. On the other hand, the SDSS measurement could be 
affected by some bias due to the limited signal-to-noise ratio at the blue
end of the spectra corresponding to $z\sim$2.
Motivated by the importance of these issues on our understanding of galaxy formation and evolution, and 
by the discrepancy in the results of different surveys, 
we developed robust fully automatic procedures to search for, detect and analyse DLAs in the seventh and last
data release (DR7) of QSO spectra from the Sloan Digital Sky Survey II. We present our methods and the algorithms used 
to analyse the data in Sections~\ref{qs} and \ref{dlas}, the statistical results on the neutral gas 
column density distribution and cosmic evolution of $\omegagdla$ in Section~\ref{res}, with a 
special emphasis on discussing systematic effects. We conclude in Section~\ref{concl}.

\section{Quasar sample and redshift path \label{qs}}

The quasar sample is drawn from the Sloan Digital Sky Survey Data Release 7 \citep{Abazajian09} and 
includes every point-source spectroscopically confirmed as quasar (specClass=QSO or HIZ\_QSO).
Basically, SDSS quasars are pre-selected either upon their colours, avoiding the stellar locus 
\citep{Newberg97}, or from matching the FIRST radio source catalogue (\citealt{Becker95}, see \citealt{Richards02} for 
a full description of the quasar selection algorithm in SDSS) and then confirmed spectroscopically.

Given the blue limit of the SDSS spectrograph (3800~\AA), we selected quasars whose emission redshift 
is larger than $\zem=2.17$, with a confidence level higher than 0.9. 
This gives 14\,616 quasar spectra that were retrieved from the 
SDSS website~\footnote{\url{http://www.sdss.org}}.

\subsection{Redshift path \label{zpath}}

The first step is to define the redshift range over which to search for DLAs along each line of sight. 
As in most DLA surveys, we define the maximum redshift, $\zmax$, at $5\,000~\kms$ bluewards of
the QSO emission redshift. 
This is mostly to avoid DLAs located in the vicinity of the quasar \citep[e.g.][]{Ellison02}.
For defining the minimum redshift, we note that 
Lyman-limit systems (LLS) prohibit the detection of any absorption feature at wavelengths 
shorter than the corresponding Lyman break ($\lambda<(1+z_{\rm LLS})\times\lambda_{912}$). 
To ensure the minimum redshift, $\zmin^0$, is set redwards of any Lyman break possibly present in 
the spectrum, we run a 2000\,\kms-wide (29 pixels) sliding window starting from the blue end of the 
spectrum and define $\zmin^0$ as the centre of the first window where the median signal-to-noise 
ratio exceeds 4.

This definition is very similar to that of \citet{Prochaska05} and guaranties 
robust detections of DLAs in spectra of minimum SNR. 
3347 quasars with $\zmin^0>\zmax$ were obviously not considered any further.
Note that the actual minimum redshift can be affected by the presence of a DLA located by chance at
the blue limit of the wavelength range. Given the importance of this effect, we postpone
its full discussion to Section~4.2.

\subsection{Quality of the spectra \label{ssnr}}

It is very difficult to control the search for DLAs in spectra of poor quality. 
Therefore, the spectra of bad quality should be removed a priori from the sample. 
For this, we must use an indicator of the spectral quality 
in the redshift range to be searched for, [$\zmino,\zmax$], that does not 
depend upon the presence of a damped \lya\ absorption, otherwise we may introduce a bias 
against DLA-bearing lines of sight.

We estimated the quality of the spectra -- independently of the presence of DLAs -- 
by measuring the median {\sl continuum}-to-noise ratio in the redshift range [$\zmino,\zmax$] defined above. 
The continuum over the Lyman-$\alpha$ forest (i.e. the unabsorbed quasar flux) is estimated by fitting 
a power law to the quasar spectrum, including the wavelength range 1215.67$\times[1+\zmino, 1+\zmax]$ in the 
blue of the \lya\ emission and regions free from emission lines in the red. We ignore the range 
5575-5585~{\AA} which is affected by the presence of dead pixels in the CCD. 
Deviant pixels, mainly due to \lya\ absorption lines in the blue and metal absorption lines in the red 
are first ignored by using Savitsky-Golay filtering. Then we iteratively remove deviant pixels 
by decreasing their weight at each iteration. This procedure converges very quickly.
A double Gaussian is then fitted on top of the \lya$+$\NV\ emission lines to reproduce 
the increased flux close to \zmax. The noise is taken from the error array.
Quasar spectra with median continuum-to-noise ratio lower than four were not considered any further. 
We are then left with 9597 quasar spectra.

\subsection{Broad absorption line (BAL) quasars \label{sbal}}

Broad absorption lines from gas associated with the quasar can possibly be confused with DLAs.
Our purpose here is not to recognise all BAL quasars but rather to automatically select 
the quasars without strong BAL outflows in order to avoid contamination of the DLA sample 
by broad lines, i.e., \OVI, \HI\ and \NV.

Therefore outflows were automatically identified by searching 
for wide absorptions \citep[extended over a few thousand kilometres per second, see][]{Weymann91} 
close to the quasar \SiIV\ and/or \CIV\ emission lines. 

\begin{figure}
\centering
\includegraphics[bb=42 399 588 755, width=\hsize]{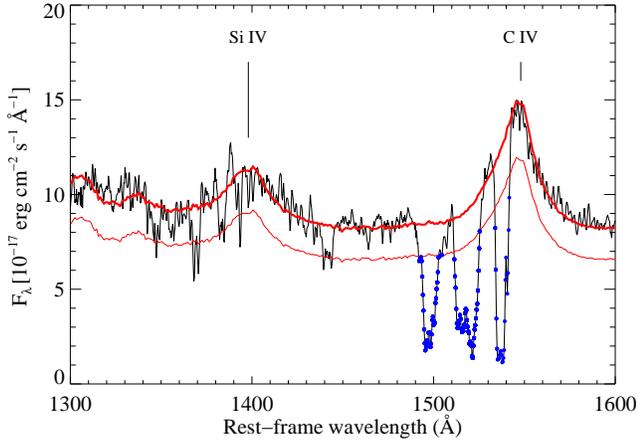}
\caption{\label{example_bal_search} Spectrum of the quasar SDSS\,J104109.86$+$001051.76 featuring 
broad absorption lines. 
The fit to the quasar continuum is shown by the thick line. The thin line represents a decrement 
of 20\% in the quasar flux. Pixels satisfying the BAL criteria inside the running windows (see text) are marked with blue dots.}
\end{figure}

For this, we computed the normalised spectrum ${\cal R}(\lambda)$ in the region
$\lambda_{\rm obs}=$[1350,1550]$\times(1+\zem)$ as the ratio of the observed spectrum
to the continuum derived following the procedure by Gibson et al. (2009). 
The quasar continuum was modelled by the product of the SDSS composite spectrum 
\citep{Vandenberk01} with a third order polynomial. This allows to reproduce well the 
combination of reddening and intrinsic shape of the quasar spectrum, as well as 
the overall shape of the emission lines with a very limited number of parameters.
However, the exact shape of the \SiIV\ and \CIV\ emission lines is accurately reproduced only when
a Gaussian is added at the position of the lines.  
We also adjusted the emission redshift by cross-correlating the reddened composite spectrum with 
the observed one in the red wings of the \SiIV\ and \CIV\ emission lines, the blue wings 
being possibly affected by BALs.
An example of continuum fitting is shown in Fig.~\ref{example_bal_search}.

We excluded the quasars whenever ${\cal R}(\lambda)$ is continuously less than 0.8 
over 1000~\kms, or ${\cal R}(\lambda)<0.8$ over at least 75\% of a 3000~\kms\ wide window,
running between 1350 and 1550~\angstrom\ in the quasar's rest frame. With this definition, we are
most sensitive to broad absorption lines with balnicity indexes (BI) larger than 1000~\kms.
The balnicity index (BI) characterises the strength of a BAL (see e.g. Gibson et al. 2009).
Note that the core of a damped \lya\ absorption will be larger than this value, so that 
possible broad \HI, \OVI\ and/or \NV\ lines from systems with BI~$<1000$~\kms\ have little 
chance to mimic a DLA.
The procedure excludes 1258 BAL quasars among the 9597 quasars left after the previous steps, 
i.e. after checking 
for adequate signal-to-noise ratio and redshift range available along the line of sight.

\begin{figure}
\centering
\includegraphics[width=\hsize,bb=75 175 515 586]{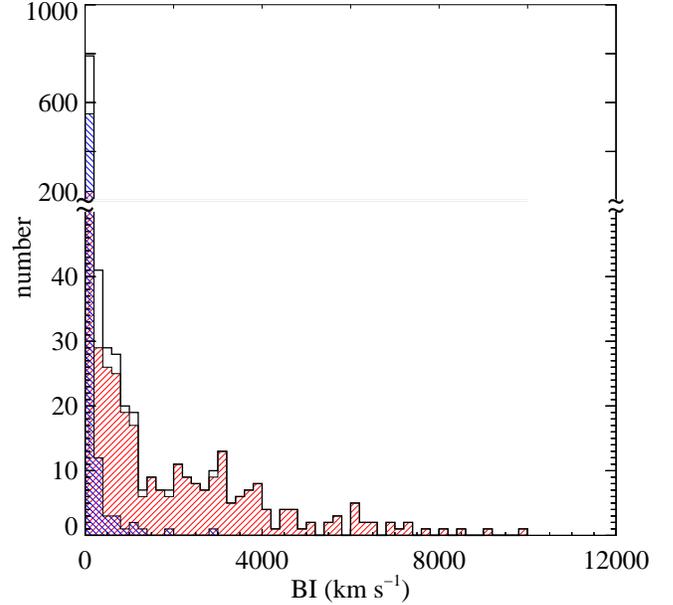}
\caption{Balnicity index (BI) distribution of the quasars in common with the BAL catalogue of 
\citet{Trump06} (solid black line histogram). The BI measurements are from these authors. The red right-dashed 
histogram corresponds to the quasars identified as BAL by our procedure, while the blue left-dashed 
histogram represents the distribution of the \citeauthor{Trump06} quasars that are 
not removed by our procedure. Very few BAL quasars with BI~$>$~500~km~s$^{-1}$ 
are missed by our automatic procedure.
\label{trump}}
\end{figure}

We have checked this automatic procedure by comparing our list of rejected quasars with the catalogue 
of SDSS BAL quasars from \citet{Trump06}. 
Fig.~\ref{trump} shows the balnicity index distribution of all quasars
in \citet{Trump06} (unfilled histogram) and that of the quasars our procedure excludes (red shaded
histogram), the difference is shown as a blue shaded histogram. It can be seen that for 
BI~$>$~500~km~s$^{-1}$ our procedure misses very few BAL QSOs. 
We have checked that the missed BAL QSOs have no strong \OVI, \HI\ or \NV\ absorption line that 
could mimic a DLA. 

We are thus left with the spectra of 8339 QSOs, without strong BALs and with sufficient signal-to-noise ratio 
to search for damped Lyman-$\alpha$ systems. We call this sample \s$_{QSO}^0$.
Note that the study of \citet[][hereafter PW09]{Prochaska09s} using DR5 is based on 7482 QSOs.

\section{Detection of DLAs and $N(\HI)$ measurements \label{dlas}}
DLA candidates are generally identified in low-resolution spectra by their large 
\lya\ equivalent widths ($W_{\rm r}\ge10$~{\AA}) and are then confirmed by higher 
spectral resolution observations \citep[e.g.][]{Wolfe95}.
However the resolving power of SDSS spectra ($R=\lambda/{\delta \lambda}\sim 1800$)
is high enough to detect the wings of damped Lyman-$\alpha$ lines. Therefore, it is possible 
not only to identify these lines but also to measure the corresponding \HI\ column 
densities from Voigt-profile fitting. 
Note that, because $R$ is constant along the spectrum, 
the pixel size is constant in velocity-space and DLA profiles of a given $N(\HI)$ are equivalently 
sampled by the same number of pixels, regardless of their redshift.

The number of candidates in the SDSS is so large that the detection 
techniques must be automatised.
\citet{Prochaska04} searched for DLA candidates in SDSS spectra by running a narrow window
along the spectra to identify the core of the DLA trough 
as a region where the signal-to-noise ratio is significantly lower than the 
characteristic SNR in the vicinity. 
Candidates were then checked by eye and the \HI\ column density measured by Voigt-profile
fitting interactively.

We develop here a novel approach which makes use of all the information available in 
the DLA profile and, most importantly, is fully automatic.  

\subsection{Search for strong absorptions \label{corrdla}}
The technique is based on a Spearman correlation analysis.
At each pixel $i$ in the spectrum, corresponding to a given redshift for Lyman-$\alpha$, 
synthetic Voigt profiles corresponding to different column densities ($N(\HI)_{\rm j}$) 
differing by 0.1~dex are successively correlated with the observed spectrum over a
velocity interval [${\rm v}_{\rm min}$,${\rm v}_{\rm max}$] corresponding to a decrement larger 
than 20\% in the Voigt profile.
Each redshift ($z_{\rm i}$) for which the Spearman's correlation 
coefficient is larger than 0.5 with high significance ($>5\sigma$) is recorded. We then add the 
criterion that the area between the observed 
($F_{\rm obs}$) and synthetic ($F_{\rm synt}$) profiles is less than the integrated error 
array ($E_{\rm obs}$) on the interval [$v_{\rm min}$,$v_{\rm max}$]:
\begin{equation}
\int_{{\rm v}_{\rm min}}^{{\rm v}_{\rm max}} (F_{\rm obs}-F_{\rm synt})^{+} 
\le \int_{{\rm v}_{\rm min}}^{{\rm v}_{\rm max}} E_{\rm obs},
\end{equation}
where $(F_{\rm obs}-F_{\rm synt})^{+}=(F_{\rm obs}-F_{\rm synt})$ if $(F_{\rm obs}>F_{\rm synt})$ and 0 otherwise.
This definition allows the DLA line to be blended with intervening Lyman-$\alpha$ absorbers.

The ($z_{\rm i}, N(\HI)_{\rm j}$) pair with highest correlation for each DLA candidate is then 
recorded. This provides a list of DLA candidates with first guesses of $N(\HI)$ and $\zabs$. 

\subsection{Fits of Lyman-$\alpha$ and metal lines \label{metals}}

\begin{figure}
\begin{center}
\begin{tabular}{ccc}
\multicolumn{3}{c}{\includegraphics[bb=260 58 541 737,clip=,angle=90,width=\hsize]{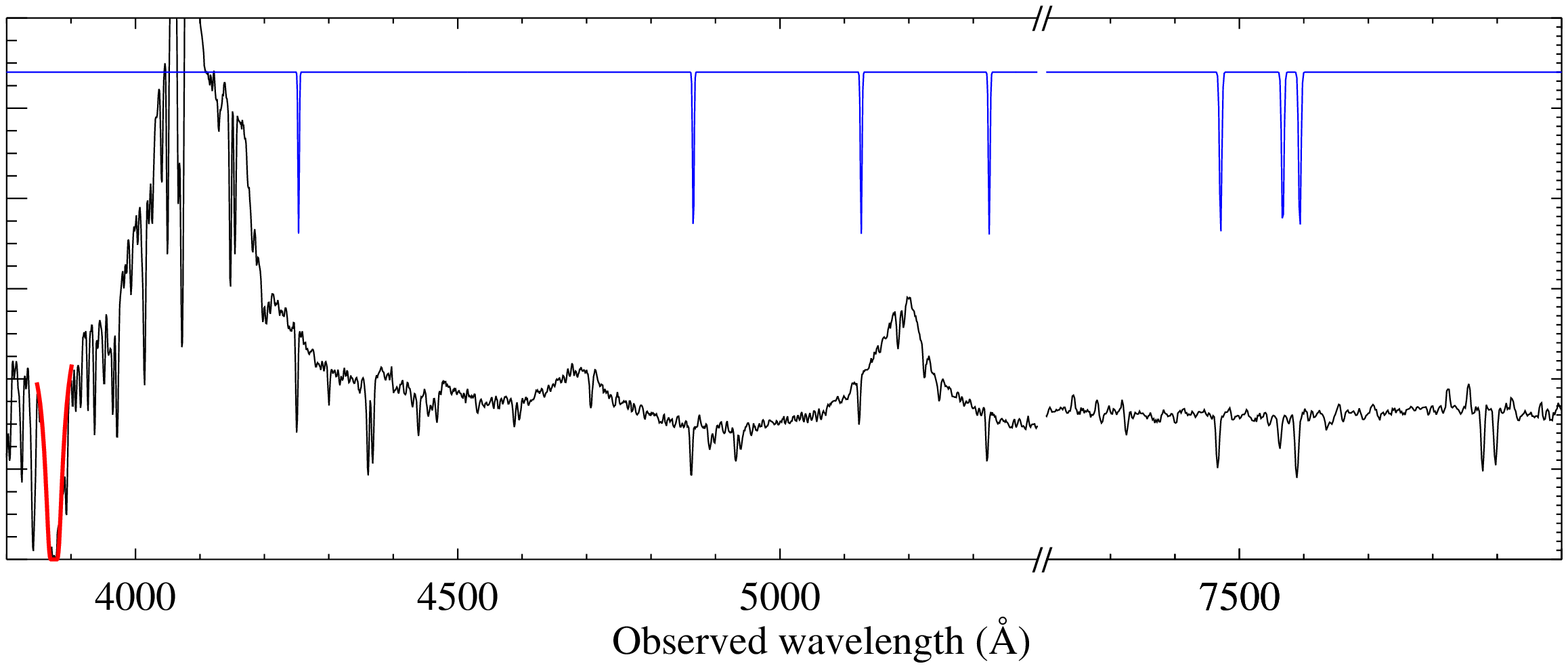}}\\
\includegraphics[bb=186 232 425 545,clip=,width=0.29\hsize]{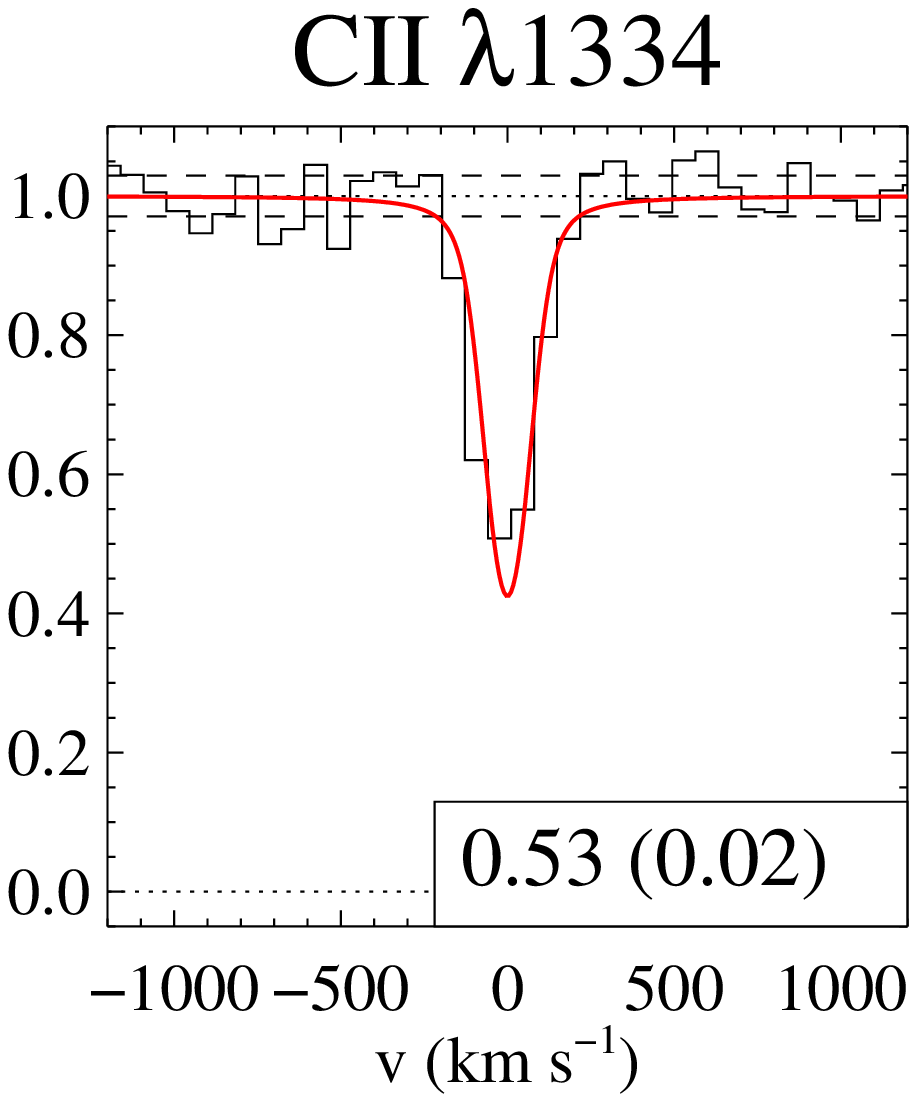} &
\includegraphics[bb=186 232 425 545,clip=,width=0.29\hsize]{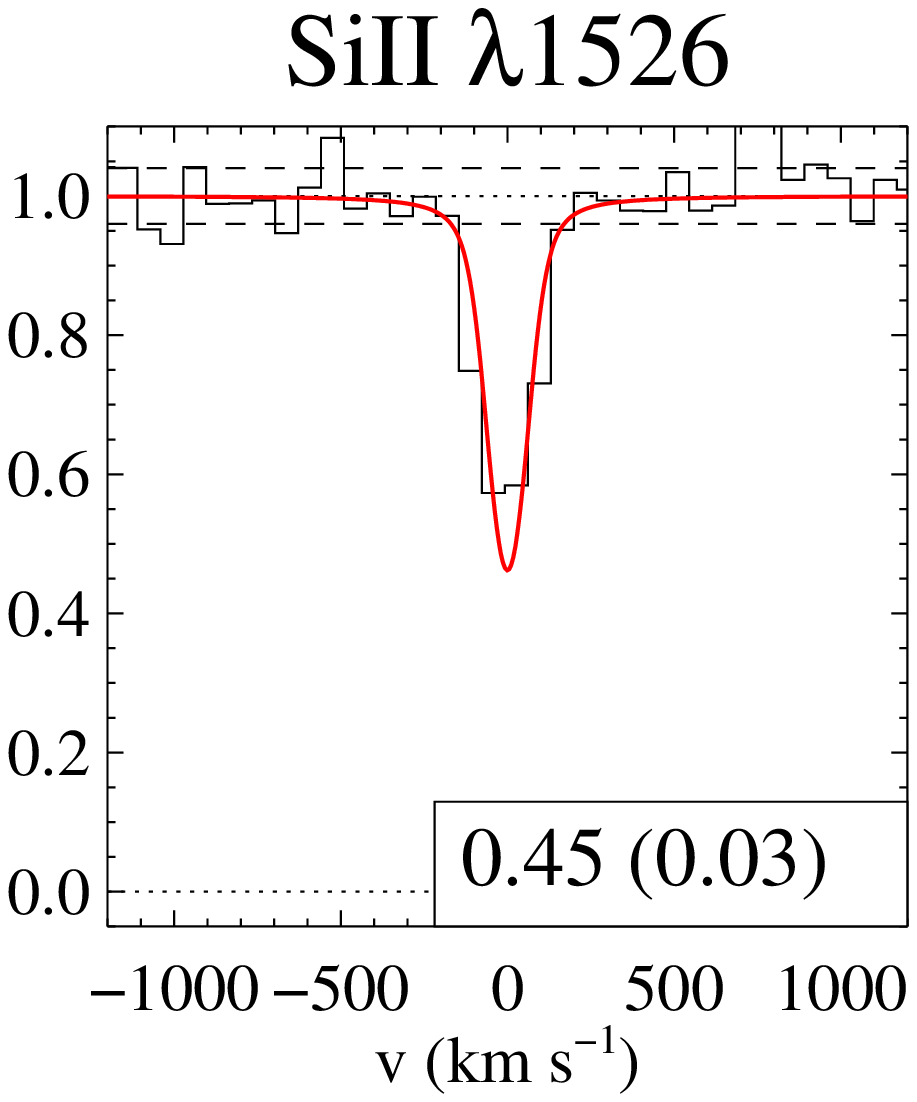} &
\includegraphics[bb=186 232 425 545,clip=,width=0.29\hsize]{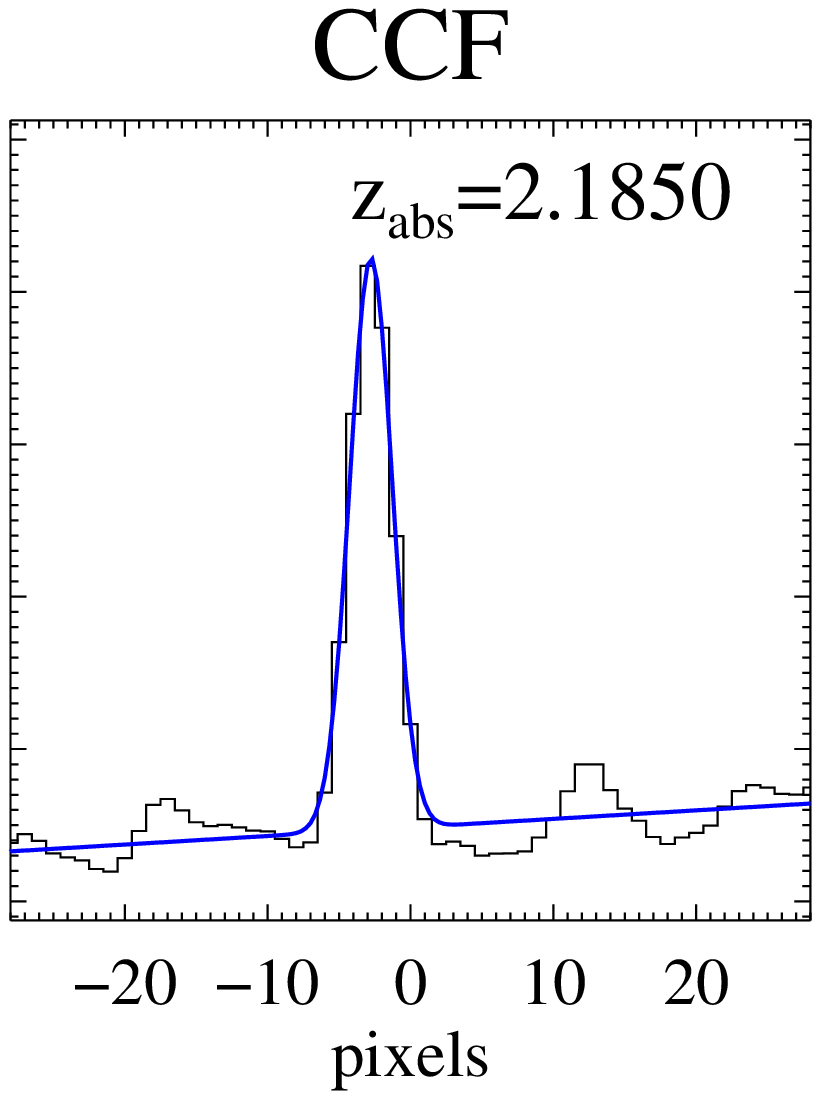} \\
\includegraphics[bb=186 232 425 545,clip=,width=0.29\hsize]{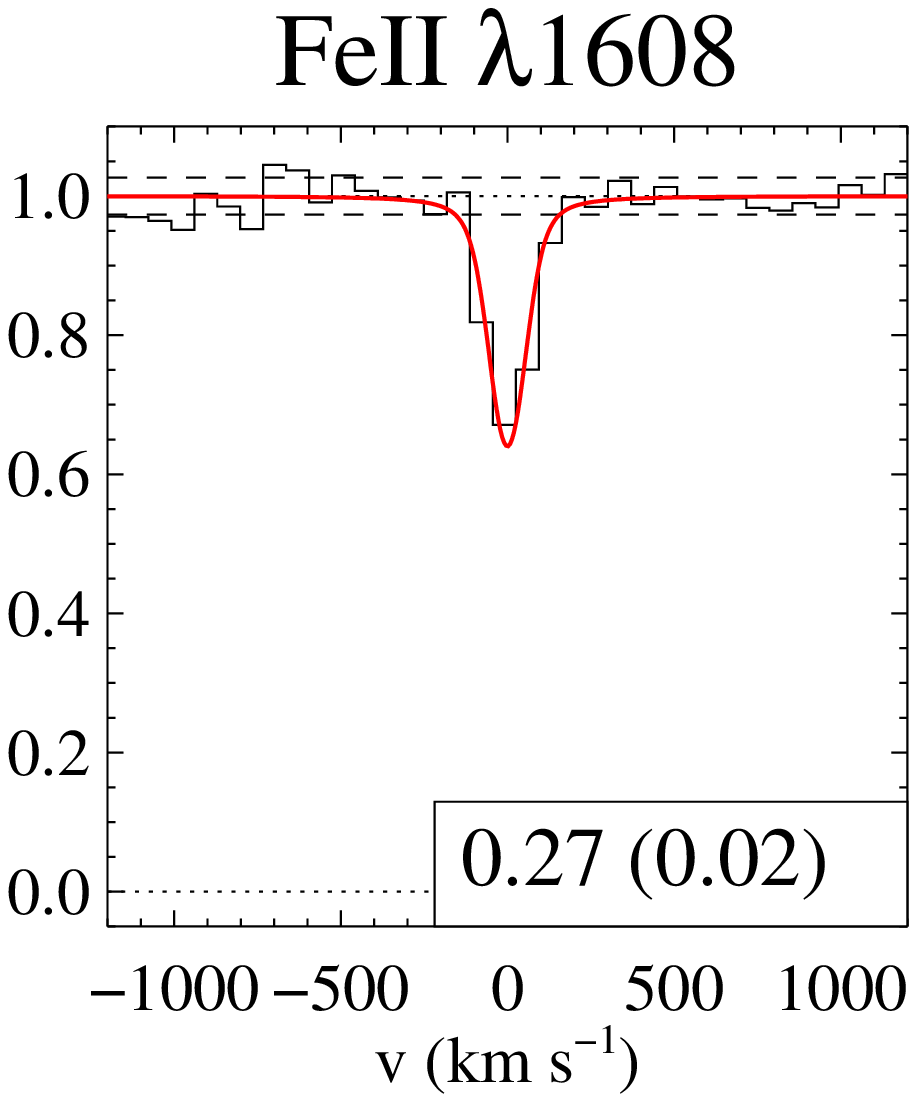} &  
\includegraphics[bb=186 232 425 545,clip=,width=0.29\hsize]{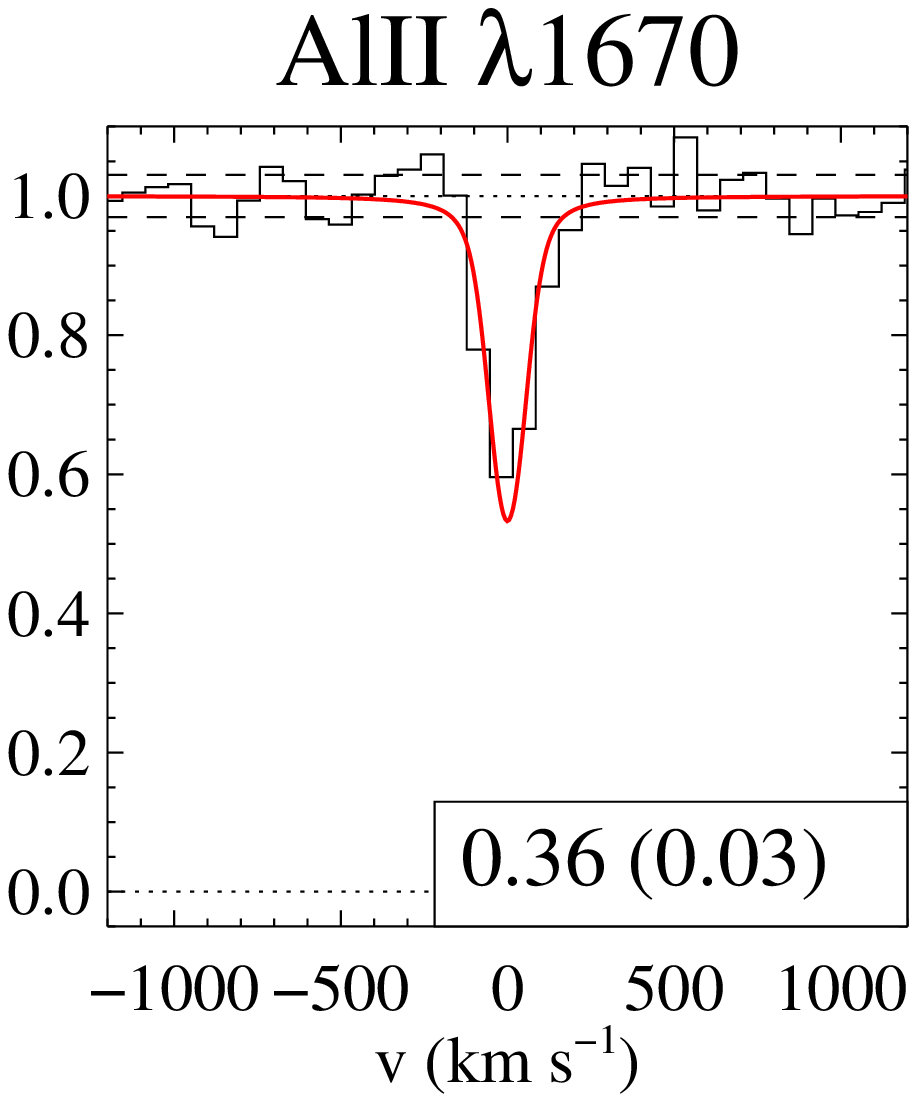} &
\includegraphics[bb=186 232 425 545,clip=,width=0.29\hsize]{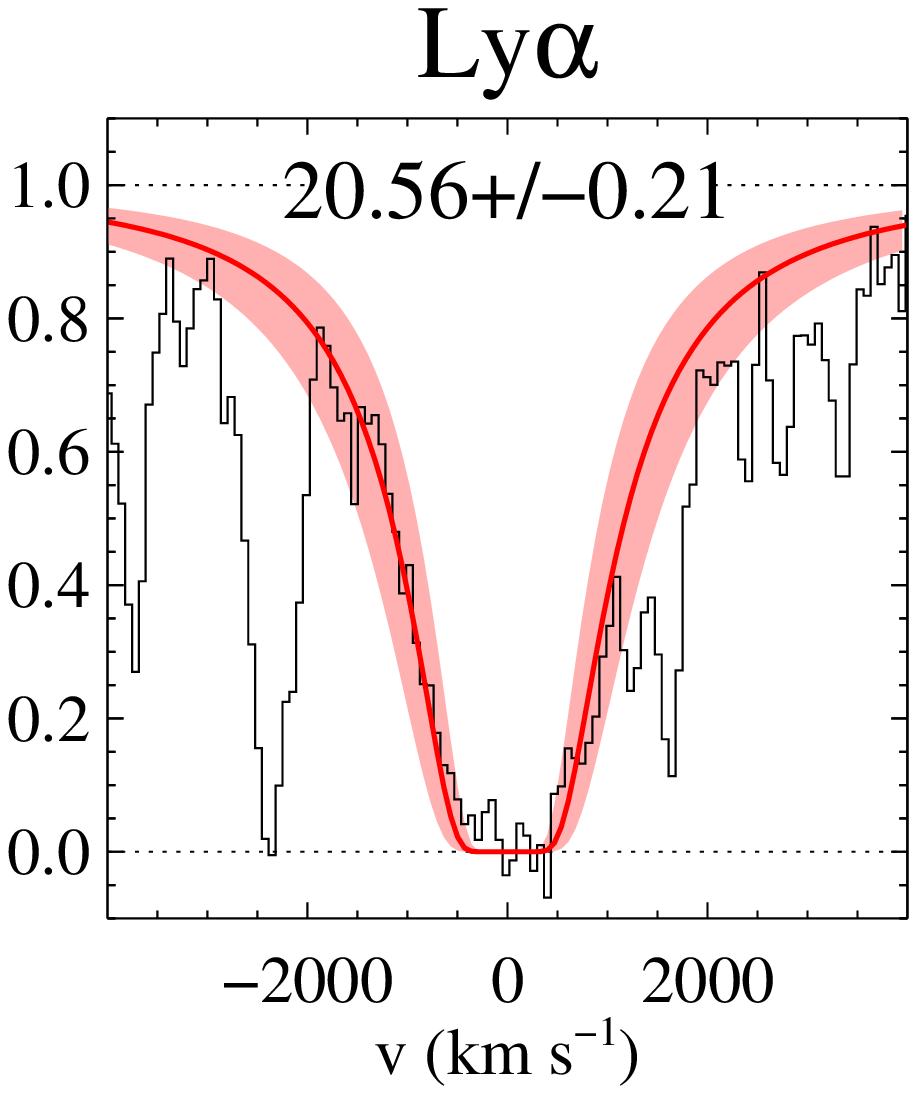} \\
\end{tabular}
\end{center}
\caption{Example of the SDSS spectrum of J\,152529.18$+$292813.18 with a DLA line at $\lambda_{\rm obs}\sim3880~{\AA}$ 
(top panel). The mask (template of low-ionisation absorption lines) used to derive an accurate 
redshift ($z_{\rm abs}$~=~2.1850) from the cross-correlation function (shown in panel CCF with 
the origin of the pixel scale set from the first guess) is overplotted in blue in the upper panel. 
The automatic fits to a few metal absorption lines are shown in the left-hand side and middle
panels (\CII\,$\lambda1334$, \SiII\,$\lambda1526$, \FeII\,$\lambda1608$, \AlII\,$\lambda1670$).
Rest-frame equivalent widths and associated errors are indicated (in \AA) at the bottom of each panel. 
The automatic Voigt-profile fit to the damped \lya\ line is overplotted to the observed spectrum 
in the bottom-right panel. The shaded area corresponds to the uncertainty on the column density
($N(\HI)=10^{20.56\pm0.21}$~\cmsq).
\label{figexample}}
\end{figure}

For each candidate, we perform, in the vicinity of the candidate redshift $z_{\rm i}$,
a cross-correlation of the observed spectrum with an absorption template 
representing the most prominent low-ionisation metal absorption lines 
(\CII$\lambda1334$, \SiII$\lambda$1526, \AlII$\lambda1670$, 
\FeII$\lambda1608,2344,2374,2382,2586,2600$, \MgII$\lambda\lambda2796,2803$).  
The template is a variant of a binary mask \citep[similar to those used to derive stellar 
radial velocities, see e.g.][]{Baranne96}, where each absorption line is represented by a 
Gaussian with a width matching the SDSS spectral 
resolution (see top panel of Fig.~\ref{figexample}). 
We restrict the mask to the metal lines expected in the red of the QSO \lya\ emission line to avoid 
contamination by \lya\ forest lines. We then obtain a sharp cross-correlation function 
(CCF, Fig.~\ref{figexample}) which is itself fitted with a Gaussian profile to derive
a measurement of the redshift with an accuracy better than $10^{-3}$.
Voigt profiles are fitted to each metal line after determination of the local continuum,
providing a measure of their equivalent widths. 
In case of non-detection of low-ionisation lines, the same procedure is repeated to detect 
\CIV\ and \SiIV\ absorption lines. 
Finally, when no metal absorption line is detected automatically, we keep the redshift 
obtained from the best correlation with the synthetic \lya\ profile ($z_{\rm j}$, see Section~\ref{corrdla}).

A Voigt-profile fit of the damped Lyman-$\alpha$ absorption is then performed to derive an 
accurate measurement of $N(\HI)$ (bottom-right panel of Fig.~\ref{figexample}), taking as initial 
value the guess on $N(\HI)$ from the best synthetic profile correlation and fixing the redshift 
to the value derived as described above. Absorption lines from the \lya\ forest are ignored 
iteratively by rejecting deviant pixels with a smaller tolerance at each iteration. 

\begin{figure}
\centering
\includegraphics[width=0.95\hsize,bb=45 245 525 497]{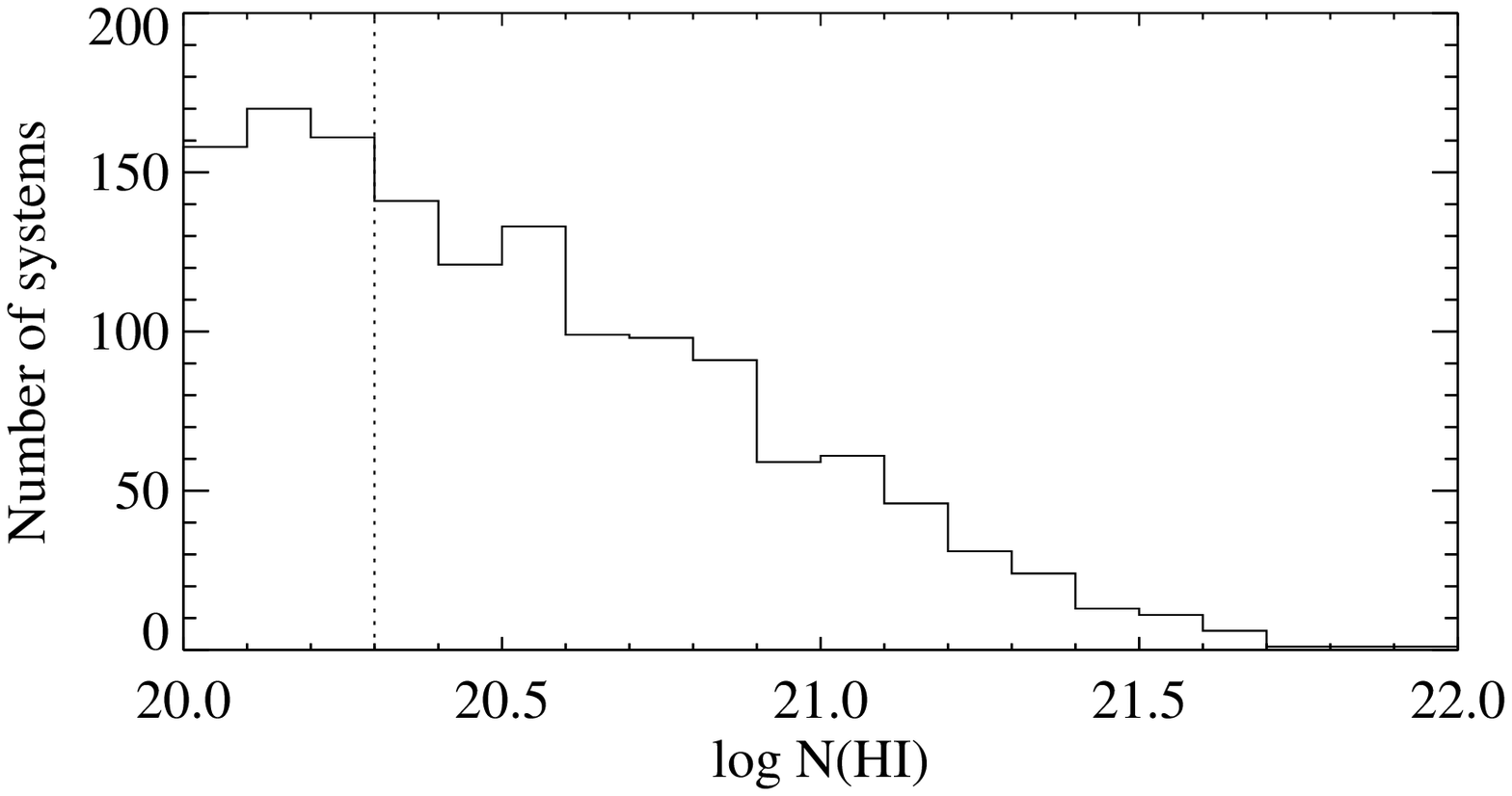}
\includegraphics[width=0.95\hsize,bb=45 245 525 497]{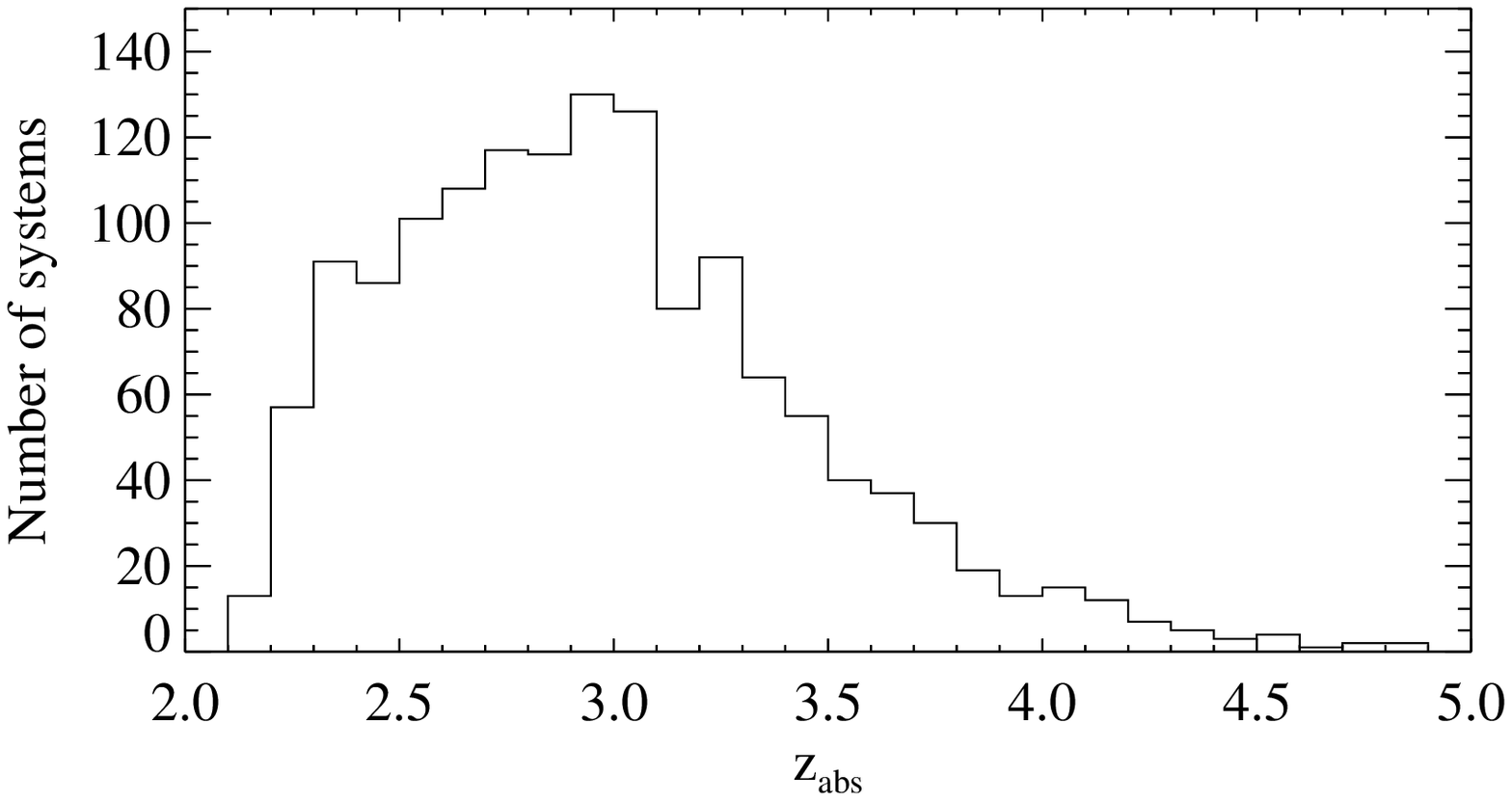}
\caption{{\sl Top}: Histogram of \HI\ column densities for the 1426 absorbers automatically detected
in SDSS DR7. The vertical dotted line marks the traditional DLA threshold value of $\log N(\HI)=20.3$. 
{\sl Bottom}: Histogram of absorption redshifts for the same sample. \label{hist_nhi}}
\end{figure}

For each system, we obtain an accurate measurement of the absorber's redshift, 
the \HI\ column density and the equivalent width of associated metal absorption lines, without any human 
intervention. We found 1426 absorbers with $\log N(\HI)\ge20$, among which 937 have $\log N(\HI)\ge 20.3$. 
This is the largest DLA sample ever built. 
The distributions of \HI\ column densities and redshifts for the whole sample is shown on Fig.~\ref{hist_nhi}.  

\subsection{Accuracy of the measurements and systematic errors \label{accuracy}}

Direct comparisons can be performed between the sample of \citetalias{Prochaska09s}, derived
from SDSS Data Release 5, and the corresponding sample from our survey.
Indeed, one would like to assess the completeness of each sample and the reliability of the 
corresponding detection procedures. 

For this, we consider only systems that are detected along lines of sight covered by both surveys.
We compare the lists of systems with $\log N(\HI)> 21$ from this work and from \citetalias{Prochaska09s}.
We check whether $\log N(\HI)> 21$ systems from a given survey are detected in the other survey, 
{\sl whatever the column density estimated in the second survey is}. The limit on $N(\HI)$ is chosen high 
enough so that no system is 
excluded because of errors. In addition, these systems are the most important for the analysis because, as we 
will show, they contribute to more than one half of $\omegagdla$. 
Only one such DLA (at $\zabs=3.755$ towards J130259.60$+$433504.5) among 76 in the 
\citetalias{Prochaska09s} list has been missed by our procedure, due to spurious lines 
leading to a wrong CCF redshift measurement. 
This means that our completeness 
at $\log N(\HI)> 21$ is about $99\%$. In turn, we discovered 6 $\log N(\HI)>21$ DLAs, with redshifts 
in the range $z=2.3-3.8$, that are not in the \citetalias{Prochaska09s} list when they should be since the 
corresponding lines of sight have been considered and the redshift of the DLA covered. These are 
J100325.13$+$325307.0 (\zabs=2.330); J205509.49$-$071748.6 (\zabs=3.553); J093251.00$+$090733.9 (\zabs=2.342),
J133042.52$-$011927.5 (\zabs=2.881); J092914.49$+$282529.1 (\zabs=2.314) and J151037.18$+$340220.6 (\zabs=2.323). Associated metal lines are detected for five of them.

The completeness at smaller $N(\HI)$ is more difficult to estimate because of the uncertainty of 
individual measurements. We compared however our detections to that of \citetalias{Prochaska09s} and 
found that we recover more than 96\% of their systems with most of the DLAs missed having 
$\log N(\HI)\sim20.3$. We also checked visually one hundred randomly-selected DLAs and found that about 
3\% of the systems in our sample are false-positive detections (1\% at $z<3.2$ and 7\% at $z>3.2$).
The completeness of both samples (\citetalias{Prochaska09s} and ours) is sufficiently high to have 
little influence on the determination of the cosmological mass density of neutral gas.  

\begin{figure}
\includegraphics[width=\hsize,bb=75 175 515 586]{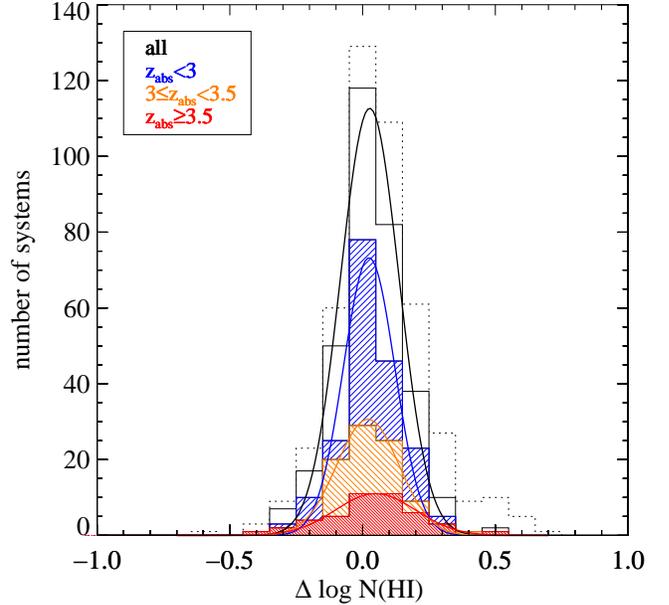}
\caption{Distribution of the difference in the column density measurements from the two surveys, 
$\Delta \log N(\HI)=\log N(\HI)_{\rm this~work} - \log N(\HI)_{\rm PW09}$, for the whole sample
(black), $z<3$ (blue), $3<z<3.5$ (orange) and $z>3.5$ (red). Distributions are corrected from the 
$\log N(\HI)=20.3$ truncating effect (see text). The non-corrected distribution for the whole sample
is represented by the dotted histogram.
\label{diffnhiz}}
\end{figure}

Next, we compare the $N(\HI$) measurements for the same systems in the two samples. Fig.~\ref{diffnhiz} shows 
the distribution of the difference in the DLA column densities,
$\Delta \log N(\HI)=\log N(\HI)_{\rm this~work} - \log N(\HI)_{\rm PW09}$, for the whole sample
(black), $z<3$ (blue), $3<z<3.5$ (orange) and $z>3.5$ (red). These distributions are corrected 
from the truncating effect at $\log N(\HI)=20.3$, i.e. from the fact that some systems may have 
$\log N(\HI)\ge20.3$ in one sample but $\log N(\HI)<20.3$ in the other sample. 
Only systems for which the opposite of $\Delta \log N(\HI)$ is also allowed are considered.
The uncorrected distribution for the whole sample is shown as a dotted line. 

The dispersion in the whole sample is about 0.20~dex and matches the typical error on individual $N(\HI)$ 
measurements. 
There is a small systematic difference that increases with redshift from less than 0.03~dex at 
$z<3$ to 0.05~dex at $z>3.5$, which is likely due to the increasingly crowded \lya\ forest at higher redshift. 

\section{Results \label{res}}

In this section, we present the statistical results from our DLA survey. 
Here, we use the standard definitions for the different statistical quantities. 

\par \noindent The absorption distance $X$ is defined as 

\begin{equation}
X(z) = \int_{0}^{z}(1+z')^2 {{\Ho} \over {H(z')}} dz',
\end{equation}

where \Ho\ is the Hubble constant and 
$H(z) = \Ho \left[{(1+z^3)\Omega_{\rm m}-(1+z)^2(\Omega_{\rm m}+\Omega_{\Lambda}-1)+\Omega_{\Lambda}}\right]^{1/2}$. 
The cosmological mass density of neutral gas, $\Omega_{\rm g}^{\HI}$, is given by

\begin{equation}
\Omega_{\rm g}^{\HI}(X) dX \equiv {{\mu m_{\rm H} \Ho} \over {c \rho_{\rm c}}} \int_{N_{\rm min}}^{N_{\rm max}} N(\HI)\fhix dX,
\end{equation}

where $\fhix$ is the $N(\HI)$ frequency distribution (i.e. $\fhix dN dX$ is the number of systems within $(N,N+dN)$ and $(X,X+dX)$), $\mu=1.3$ is the mean molecular mass of the gas and $\rho_{\rm c}$ is the critical mass density. 
Setting $N_{\rm min}=2\times 10^{20}$~\cmsq~ and $N_{\rm max}=\infty$ gives $\omegagdla$, the cosmological mass density of neutral gas in DLAs. Since at the column densities of DLAs the gas is neutral, this is also the total mass density 
of the gas in DLAs. In the discrete limit, $\omegagdla$ is given by:

\begin{equation}
\omegagdla = {{\mu m_{\rm H} \Ho} \over {c \rho_{c}}} {{\Sigma N(\HI)} \over {\Delta X}} ,
\end{equation}

where the sum is calculated for systems with $\log N(\HI)\ge 20.3$ along lines of sight with a 
total pathlength $\Delta X$.
We adopt a $\Lambda$CDM cosmology with $\Omega_{\Lambda}=0.7$, $\Omega_{\rm m}=0.3$, and $\Ho=70$~\kms\,Mpc$^{-1}$ 
\citep[e.g.][]{Spergel03}.

\subsection{Sensitivity function}

\begin{figure}
\centering
\includegraphics[width=\hsize,bb=75 175 515 586]{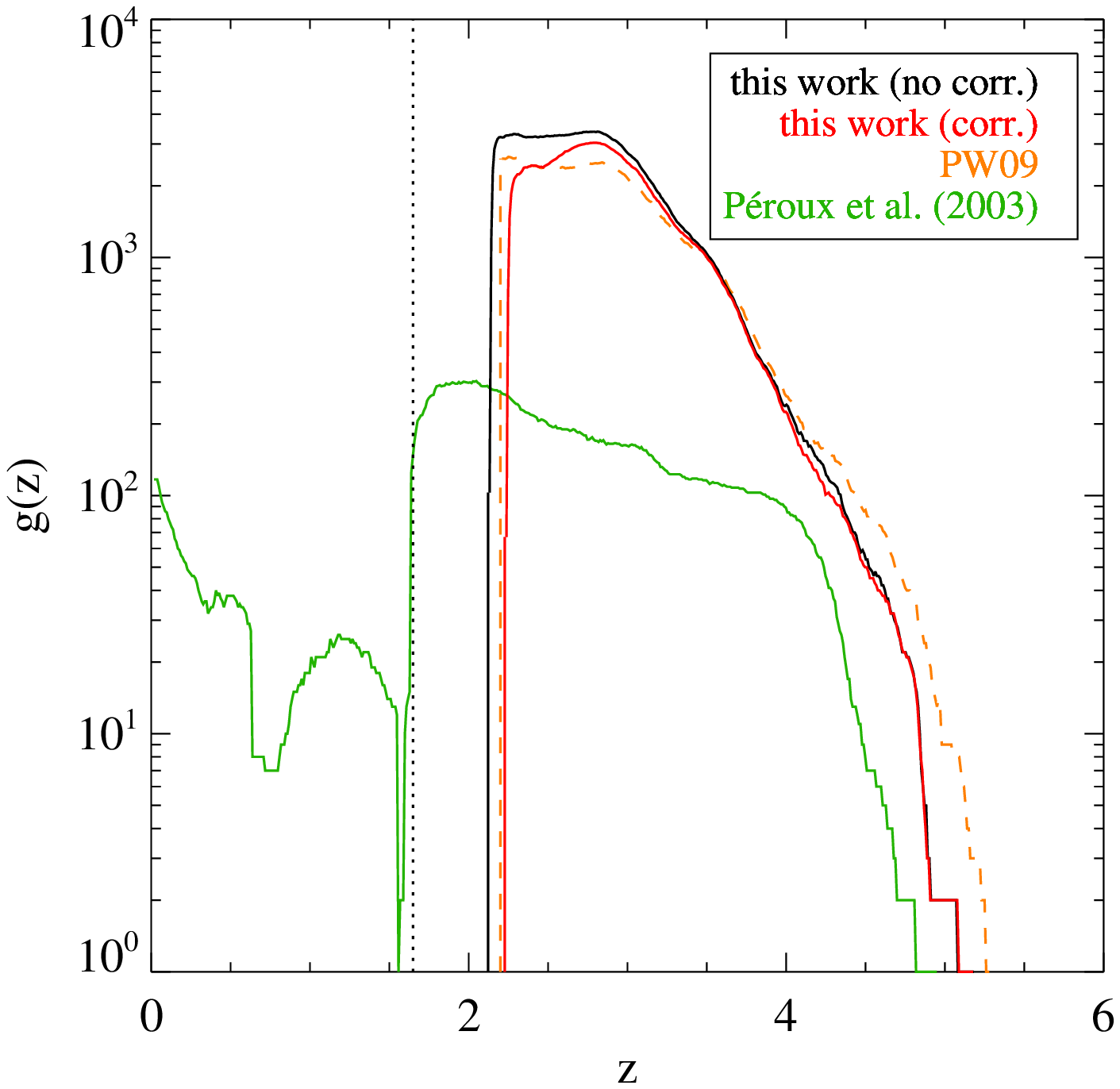}
\caption{Redshift sensitivity function $\gz$ of the different DLA surveys considered in this paper. The black and 
red curves represent respectively the sensitivity of our survey without ($\delta {\rm v}=0$~\kms) and with 
($\delta {\rm v}=10\,000$~\kms) correction of the edge-effect bias (see Sect.~\ref{sbias}). The green curve is from the 
sample built by \citet{Peroux03} whereas the orange dashed curve represents the DR5 sensitivity 
from \citet{Prochaska09s}.
The vertical dotted line corresponds
to $z=1.65$, which is the redshift below which the \lya\ line cannot be observed from the ground because of
the atmospheric absorption. 
\label{gz}}
\end{figure}

We present in Fig.~\ref{gz} the sensitivity functions $\gz$, i.e. the number of lines of sight covering a 
given redshift, for the different DLA surveys discussed in this paper. 

The redshift sensitivity of SDSS is significantly 
larger than that of the largest QSO compilation prior to SDSS \citep{Peroux03} at any redshift larger than 2.2. 
The higher sensitivity ($\sim30\%$) of our SDSS sample compared to that of \citetalias{Prochaska09s} 
over the redshift range $z\sim2.2-3.5$ reflects the increase in the number of observed quasars between 
the two data releases (DR5 and DR7). At $z>3.5$, the 
sensitivity of the \citetalias{Prochaska09s} quasar sample is higher than that presented here. 
This is due to (i) the choice by these authors to exclude only 3000~km~s$^{-1}$ from the emission
redshift of the quasar whereas we exclude 5000~km~s$^{-1}$ to avoid the proximity effect, 
(ii) the slightly strongest constraint on SNR we impose when including quasar spectra in our survey, 
(iii) the inclusion of quasars with some BAL activity in the \citetalias{Prochaska09s} sample and 
(iv) the fact that we restrict our quasar sample to those with confidence on the redshift measurement 
higher than 0.9.
Very few quasars are detected in SDSS at $z>4$ and the determination of 
$\omegagdla$ at these redshifts would benefit of dedicated surveys \citep[see][]{Guimaraes09}. 
We note that among the ten $\zem\ge5$ quasars in the \citetalias{Prochaska09s} sample, 
J\,165902.12$+$270935.1 was spectroscopically mis-classified as `galaxy' instead of `QSO' by the SDSS 
while the redshift confidence for J\,075618.13$+$410408.5 is smaller than 0.9. These lines of sight
were therefore not included in our quasar sample (see Sect.~\ref{qs}). Furthermore, five of these quasars 
have been rejected from our statistical sample, either because of BAL activity (Sect.~\ref{sbal}) or 
because of a low mean signal-to-noise ratio of the spectrum (Sect.~\ref{ssnr}).

\subsection{Importance of systematic effects \label{sbias}}

With the large number of quasar spectra available in SDSS, we reach a level 
at which systematic effects become more important than statistical errors. 
In particular, the statistical results of the survey are very sensitive to the
determination of the total absorption distance $\Delta X$.

\begin{figure}
\centering
\includegraphics[bb=400 400 585 765,angle=90,width=\hsize]{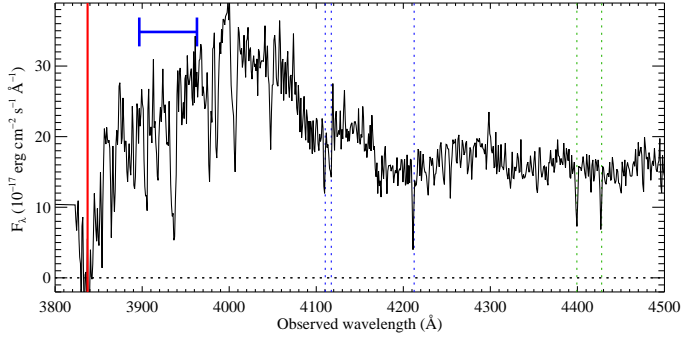}
\caption{Example of the effect of the presence of a DLA on the determination of $\zmin$. 
Because of the DLA absorption near the blue end of the spectrum (thick vertical line), 
the minimum redshift along this line of sight is set by the automatic procedure redwards of the absorption
(and the redshift range considered is marked by an horizontal segment). If the DLA had no effect on 
the determination of the minimum accessible redshift, then the latter would have been set to the blue end 
of the spectrum.
The consequence is that the DLA is missed, while it should have been included in the sample (see Sect.~\ref{sbias}).
Note that the reality of the DLA is 
confirmed by the detection of metal lines whose positions are indicated by vertical dotted lines. The spectrum 
shown in this figure is that of J162131.46$+$234550.8. We propose a procedure to avoid 
this systematic effect (see Text).
\label{missed_dv}}
\end{figure}

The presence of a DLA absorption line significantly attenuates the quasar flux and decreases
the signal-to-noise ratio of the spectrum. 
Therefore, if a DLA line is present at the blue end of the spectrum,
the minimum redshift considered along the line of sight prior to any search
for DLA absorption can be overestimated. The corresponding redshift range is rejected
a priori because of the presence of the DLA. 
The immediate consequence is that the presence of a strong absorption can preclude 
its inclusion in the statistical sample (see Fig.~\ref{missed_dv}).
We expect this effect to be important for large $N(\HI)$, when $\lambda_{\rm DLA}$ is close 
to the blue-end of the spectrum (at 3800~\angstrom), and when the signal-to-noise ratio of 
the spectrum is low. Note that the bias we describe here can affect {\sl all} DLA surveys.

\begin{figure}
\centering
\includegraphics[width=\hsize,bb=75 175 515 586,clip=]{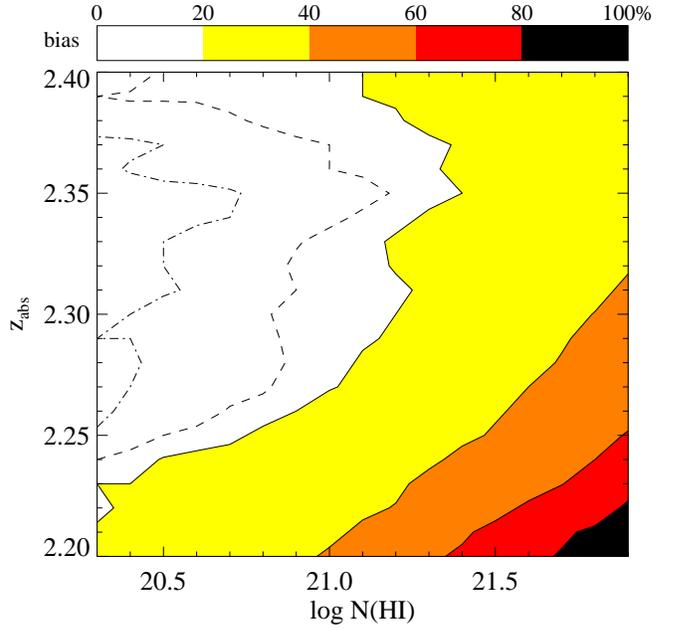} 
\caption{Result of a simulation to estimate the severity of the bias (defined as the 
percentage of DLAs missed; see Text) 
due to the incorrect pathlength determination resulting from the presence of a DLA at the blue end of the spectrum
as a function of the redshift and neutral hydrogen column density of the absorber. The colour scale 
is given at the top of the figure. Additionally, the dotted, dashed-dotted and dashed lines 
represent the 5, 10 and 15\% contours, respectively.\label{bias}}
\end{figure}

In order to assess the severity of this effect, we artificially added damped \lya\ absorptions to the spectra of 
quasars from sample \s$_{QSO}^0$. Column densities are in the range $\log N(\HI)=20.3-22$ and redshifts 
in the range $2.2 \le z \le 2.4$. 
We applied our automatic procedure to define $z_{\rm min}$ and 
compare the minimum redshift obtained along each line of sight with 
($\zmin^{0,DLA}$) and without ($\zmin^0$) the DLA. 
For each set of values ($z_{\rm abs}$, $\log N(\HI)$), we calculated 
the fraction of DLAs that are missed because of their influence on the redshift path 
($\zmin^{0,DLA}>\zabs$ while $\zmin^0<\zabs$). This fraction indicates the severity of the bias. 
The result of this exercise is shown on Fig.~\ref{bias}. It is clear from this figure that 
there is indeed a severe bias against the presence of DLAs at the blue end of the spectra. 
This bias increases, as expected, with decreasing redshift and increasing column density. 

We propose to circumvent this problem by adding a systematic velocity shift $\delta v$ 
to $\zmin^0$ defining \zmin\ as:

\begin{equation}
\label{eqcorr}
\zmin=\zmin^0+{{\delta {\rm v}}\over{c}} (1+\zmin^0).
\end{equation}

We therefore a priori exclude part of the spectrum at the blue end. This decreases
the total pathlength of the survey but guarantees, 
for a sufficiently large $\delta v$, that any pathlength that would be
excluded in the case of the presence of a DLA is a priori excluded from
the statistics. 
In other words, the definition of $[\zmin-\zmax]$ will not depend upon the 
presence of a DLA in this redshift range and the bias will be avoided {\sl a priori}. 

Note that in their survey, \citet{Prochaska05} are aware of this effect and apply a 
shift of $\delta {\rm v}=1\,500$~\kms\ 
which they claim is sufficient to avoid the bias considered as minor. They also restrict
their analysis to $z>2.2$ so that the very blue end of the spectrum is not considered.
However the wings of a DLA can significantly lower the signal-to-noise ratio over a velocity 
range as large as several thousands kilometres per second. For example, a $\log N(\HI)\sim 21$ DLA lowers 
the SNR by more than 10\%  (and obviously up to 100\% in the core of the profile) over a velocity range 
of about 10\,000~\kms. We therefore expect the bias to be corrected only for large values of $\delta {\rm v}$.

\begin{figure}
\centering
\includegraphics[width=\hsize,bb=75 175 515 586,clip=]{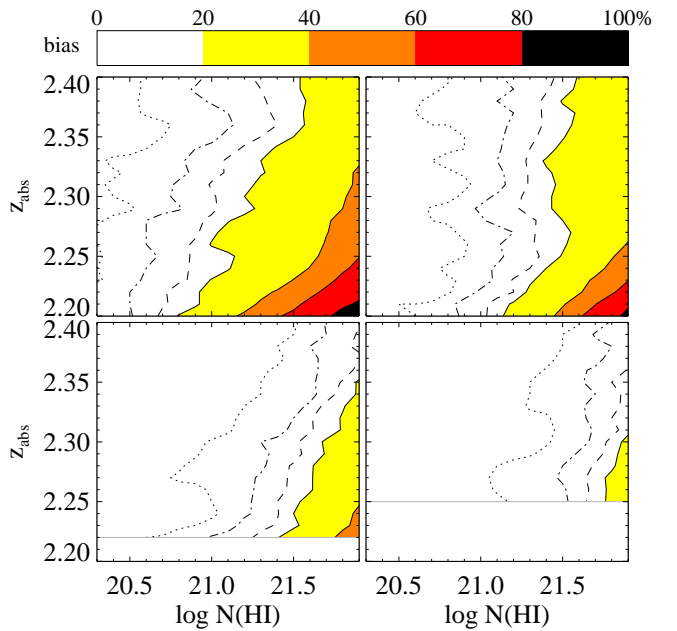}
\caption{Same as Fig.~\ref{bias} when applying different corrections. From left to right and top to bottom: 
$\delta {\rm v}=2\,500, 5\,000, 7\,500$ and $10\,000~\kms$. Colours and line styles are as per Fig.~\ref{bias}. \label{bias_corr}}
\end{figure}

We performed the same test as described above with $\delta {\rm v}=2\,500, 5\,000, 7\,500$ 
and  $10\,000$~\kms. The different
panels of Fig.~\ref{bias_corr} give the results. It is clear that the bias is 
still quite strong for $\delta {\rm v}=2\,500$~\kms\ and almost vanishes for $\delta {\rm v}=10\,000$~\kms. 

We will see in the following that the systems with $\log N(\HI)\sim 21.3$ contribute 
most to the total cosmological mass density of neutral gas. 
The residual bias must therefore be well below 10\% for this kind of column densities.
This is the case with $\delta {\rm v}=10\,000$~\kms: only systems with very large column densities 
($\log N(\HI)>21.9$) and redshifts below 2.3 have $>$~20\% probability to be missed. 
As we expect about one such system in the whole SDSS survey and at any redshift, the remaining 
bias is well below the Poissonian statistical error. Applying a $\delta {\rm v}$ shift larger than 
$10\,000~\kms$ would therefore be useless and would unnecessarily decrease the total pathlength of 
the survey (see Fig.~\ref{gz}).

\begin{figure}
\centering
\begin{tabular}{cc}
\includegraphics[bb=425 532 583 747,clip=,angle=90,width=0.47\hsize]{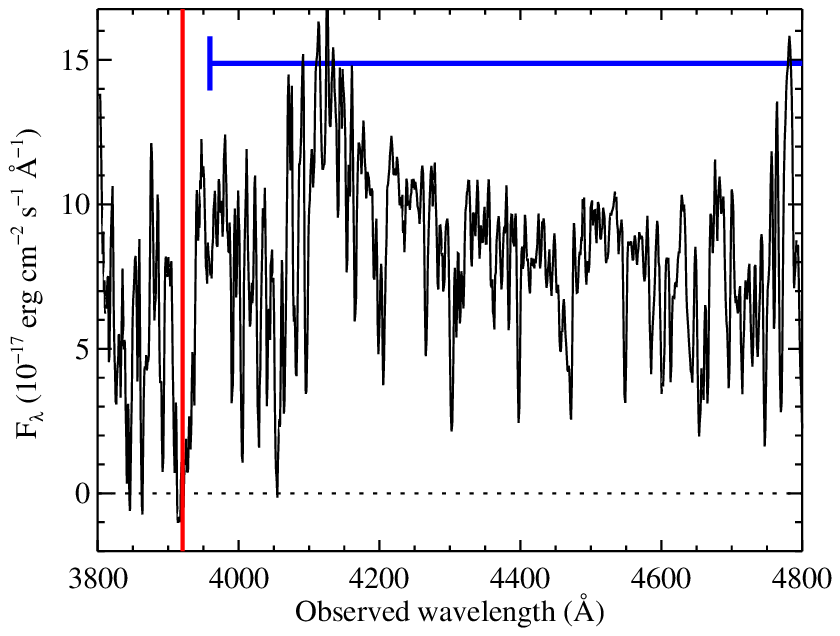} &
\includegraphics[bb=425 532 583 747,clip=,angle=90,width=0.47\hsize]{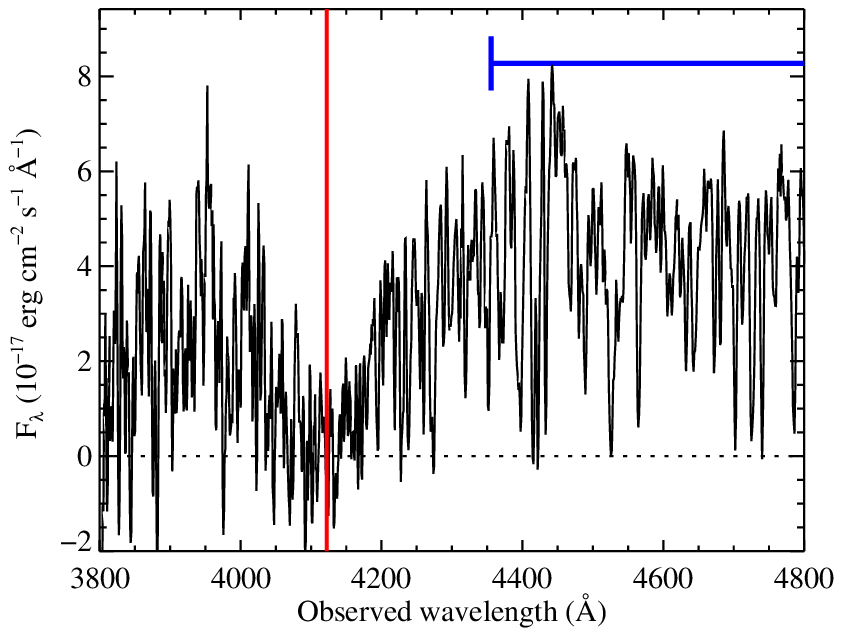} \\
\includegraphics[bb=400 532 583 747,clip=,angle=90,width=0.47\hsize]{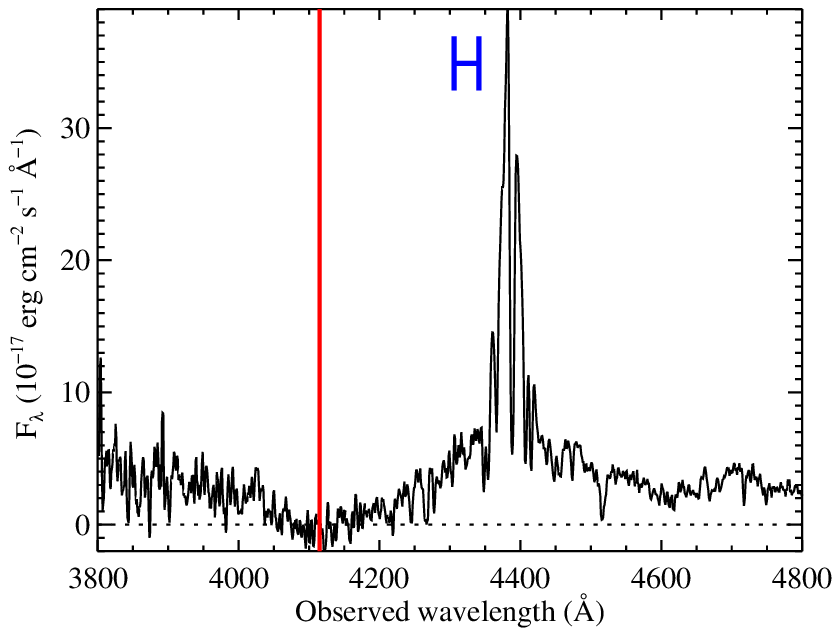} &
\includegraphics[bb=400 532 583 747,clip=,angle=90,width=0.47\hsize]{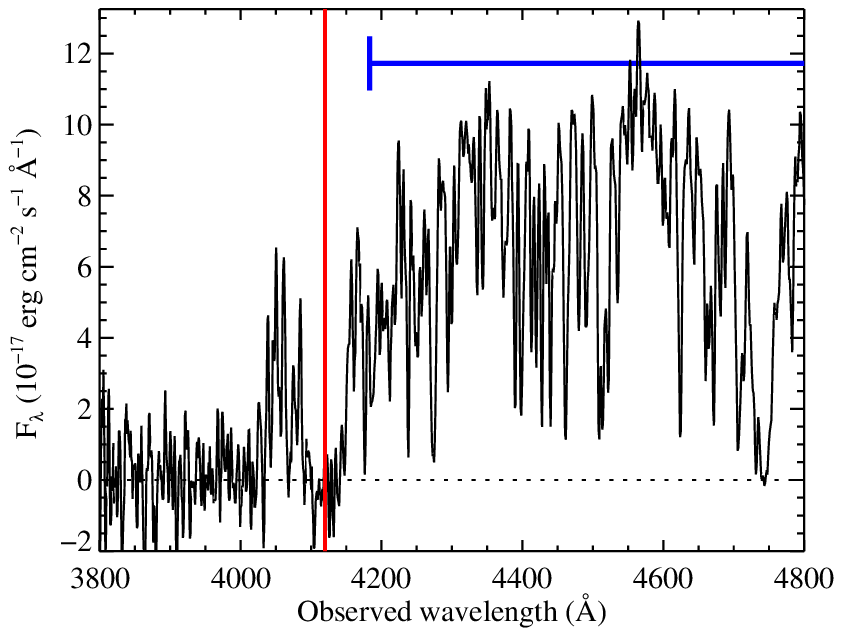} \\
\end{tabular}
\caption{Spectra of four DLAs with $\zabs<\zmin$ in the sample of \citetalias{Prochaska09s}. 
The vertical lines show the positions of the \lya\ absorptions and the horizontal segment
corresponds to the redshift pathlength (as defined 
by these authors) probed along each line of sight. In all cases, the presence of metal lines confirm 
the DLA. From left to right and top to bottom: J092322.86$+$033821.5, J155556.90$+$480015.0, J084006.65$+$362531.6 and 
J095604.44$+$344415.5. \label{missedpk}}
\end{figure}

In order to verify that the bias indeed affects the results from \citetalias{Prochaska09s}, we 
searched their quasar sample for DLAs at $z=2.2-2.4$ that their procedure missed. 
To flag these systems, we could have searched for strong \MgII\ absorption 
lines that have been shown to be good tracers of DLAs \citep[e.g.][]{Rao00}, but 
the corresponding absorption lines are unfortunately 
redshifted either beyond the SDSS spectrum ($z>2.28$) or at its very red end, where the quality of the 
spectra becomes very poor.  We therefore automatically searched for strong \FeII\ absorption lines instead.
Out of 57 systems detected this way, 32 are in \citeauthor{Prochaska09s}'s statistical 
sample and 25 have been missed because the minimum redshift is set redwards of the 
DLA due to the decreased signal-to-noise ratio.
Few examples of such DLAs are given in Fig.~\ref{missedpk}. 

In the following, we will apply the $\delta v$ cut to avoid the bias (850 bona-fide DLAs are 
then left in the statistical sample) and correct our statistical results for the
reliability of our DLA sample ($\sim 93\%$ at $z>3.2$). Note that, 
while the first correction will have important consequences, 
the second correction only has a minor effect on the $\omegagdla(z)$ results. 

\subsection{Frequency distribution}

\begin{figure}
\includegraphics[width=\hsize,bb=75 175 515 586]{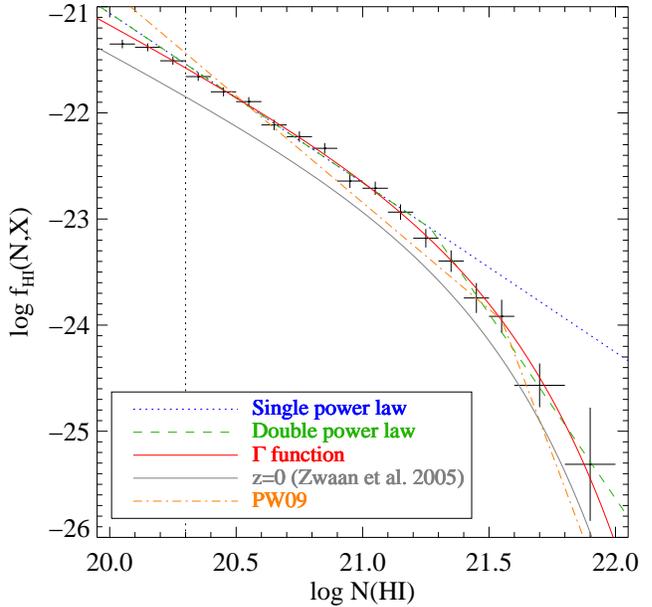}
\caption{\label{fhix} $N(\HI)$ frequency distribution of damped Lyman-$\alpha$ systems in SDSS-DR7 
from our automatic procedure. Fits to the observations by a single power law, a double power law 
and a gamma function are given as, respectively, a dotted blue, 
dashed green and solid red line. The double-power law fit to the \citetalias{Prochaska09s} 
sample is indicated by the dashed orange line. The $\Gamma$-function fit to the frequency distribution 
obtained by \citet{Zwaan05} from 21-cm observations at $z=0$ is also indicated as a solid grey line 
for direct comparison.}
\end{figure}

In Fig.~\ref{fhix}, we present the $N(\HI)$ frequency distribution function $\fhix$ of the whole sample.
Vertical error bars are representative of Poissonian statistical errors while horizontal bars 
represent the $\log N(\HI)$-binning (by steps of 0.1~dex). We find that 
a double power-law  \citepalias[e.g.][]{Prochaska09s},

\begin{equation}
\fhix = \left\{
    \begin{array}{ll}
        k_{\rm d}\left({N \over N_{\rm d}}\right)^{\alpha_{\rm d1}} & \mbox{for }  N <   N_{\rm d}\\
                                                  &                        \\
        k_{\rm d}\left({N \over N_{\rm d}}\right)^{\alpha_{\rm d2}} & \mbox{for }  N \ge N_{\rm d}\\        
    \end{array}
\right.
\end{equation}

\noindent or a $\Gamma$-function \citep[e.g.][]{Fall93,Peroux03},

\begin{equation}
\fhix=k_g\left({N \over N_g}\right)^{\alpha_g} e^{(-N/N_g)}
\end{equation}

\noindent fit the data equally well ($\chi^2_{\nu}=1.1$ and 0.7, respectively). 
The best fit values of the parameters are summarised in Table~\ref{fits_fhix}.
Slight differences between the $\Gamma$-function and double power-law fits 
(see also next Section) are not statistically significant.

\begin{table}
\centering
\caption{Parameters of the fits to the $N(\HI)$ frequency distribution (see Fig.~\ref{fhix}). \label{fits_fhix}}
\begin{tabular}{lcr rcr}
\hline
\hline
\multicolumn{3}{c}{Double power law}    & \multicolumn{3}{c}{$\Gamma$ function} \\ 
\hline

 $k_{\rm d}$      &$=$&$-23.09$       & $k_{\rm g}$     &$=$&$-22.75$ \\  
 $N_{\rm d}$      &$=$&$21.27$        & $N_{\rm g}$     &$=$&$21.26$  \\  
 $\alpha_{\rm d1}$&$=$&$-1.60$        & $\alpha_{\rm g}$&$=$&$-1.27$  \\  
 $\alpha_{\rm d2}$&$=$&$-3.48$        &                &    &        \\

\hline
\end{tabular}
\end{table}

The slope of the distribution is found to be $\alpha\sim-1.6$ for $N(\HI)<21.4$, which is close to what 
is expected from models of photo-ionised gas in hydrostatic equilibrium \citep[e.g.][]{Petitjean92}. 
This is flatter than what is found by \citet{Prochaska05}, $\alpha\sim-2$ for their whole sample
\citepalias[see also][]{Prochaska09s}. \citet{Peroux05} already mentioned that the slope of the frequency 
distribution at $N(\HI)$ around the conventional DLA threshold is flatter than $-2$. 

\begin{figure}
\includegraphics[bb=322 30 575 755, angle=90,width=\hsize]{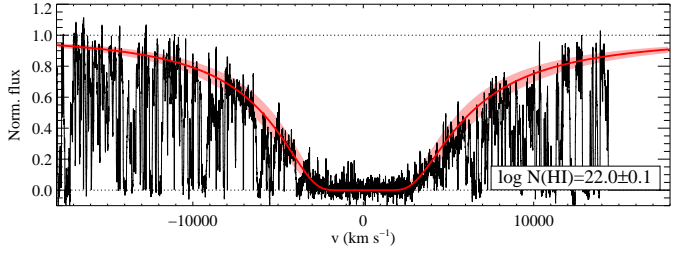}
\caption{Damped Lyman-$\alpha$ absorption line at $\zabs=3.286$ towards SDSS\,J081634.40$+$144612.86. 
The neutral hydrogen column density, measured from the Voigt profile fit to the smoothed 
(7 pixel boxcar) UVES spectrum, is $\log N(\HI)=22.0\pm0.1$.  
This is the highest \HI\ column density ever measured along QSO lines of sight.
\label{bigdla}}
\end{figure}

The slope of $\fhix$ at large $N(\HI)$, $\alpha\approx-3.5$, implies that systems with very large 
column density are very rare. We find a slope much flatter than \citetalias{Prochaska09s} ($\alpha\approx-6$).  
This is probably due to our definite detection of the first DLA with $\log N(\HI)=22$, at $\zabs=3.286$ 
towards SDSS\,J081634$+$144612.  We obtained high spectral resolution data for this quasar with UVES in April 2008. 
The column density, measured from the UVES spectrum, is $\log N(\HI)=22.0\pm0.1$ (see Fig.~\ref{bigdla}) 
while the column density derived automatically from the SDSS spectrum is $\log N(\HI)=21.92\pm0.19$.
This is the absorber with the largest column density observed to date along a quasar line of sight.
Such a column density is similar to that of DLAs detected at the redshift of Gamma-Ray Bursts 
\citep[e.g.][]{Vreeswijk04,Jakobsson06,Ledoux09}.
Detailed analysis of this system will be presented in a future paper. Note that from the fit of the frequency 
distribution, no more than one system like this one is expected in the whole SDSS survey. 

The shape of the $N(\HI)$ frequency distribution at high column densities suggests
that there is no abrupt transition between neutral hydrogen and molecular 
hydrogen in diffuse clouds as advocated by \citep{Schaye01}. This is supported by the results of 
\citet{Zwaan06} who used CO emission maps to show that the H$_2$ distribution 
function is a continuous extension of $\fhix$ for high column densities. 
In addition, there is only a small tendency for increasing H$_2$ molecular fraction with larger
$\HI$ column density \citep{Ledoux03,Noterdaeme08}.
  
\subsection{Convergence of $\omegagdla$ and the contribution of sub-damped 
Lyman-$\alpha$ systems}

One major issue when measuring the cosmological mass density of neutral gas in DLAs 
is the convergence of $\omegagdla$ at large $\log N(\HI)$ values.
An artificial cut at large $N(\HI)$ was frequently introduced to prevent the 
integration of a single power-law to diverge. This was justified by 
small number statistics at the highest column densities.
The SDSS allowed \citetalias{Prochaska09s} to observe for the first
time that the slope of $\fhix$ is steeper than $-2$ for 
$\log N(\HI)>21.5$, directly demonstrating that $\omegagdla$ converges. 

\begin{figure}
\includegraphics[width=\hsize,bb=75 175 515 586]{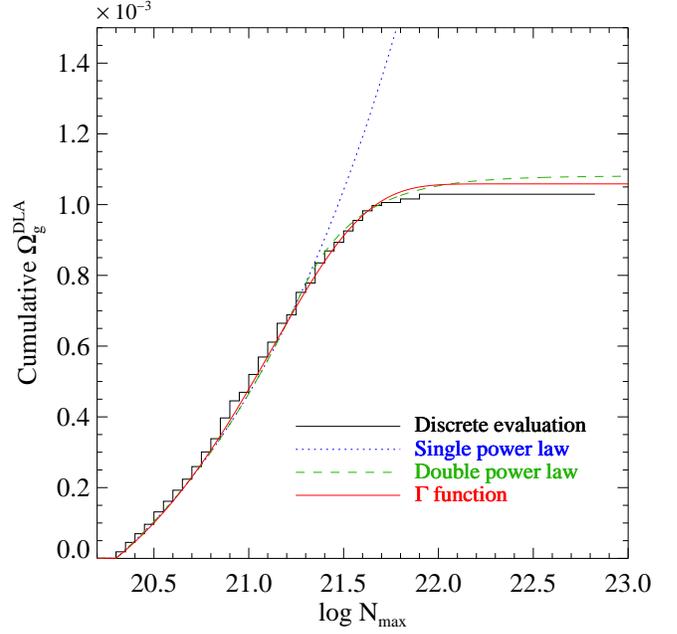}
\caption{Cumulative cosmological mass density of neutral gas in DLAs as a function of maximum column density. 
The apparent flattening of the curve at $\log N(\HI)\sim 21.7$ implies convergence. Double power-law and 
$\Gamma$ functions are equally possible solutions while a single power-law diverges and is not
representative of the data at high column densities.
\label{omegacumul}}
\end{figure}

Figure~\ref{omegacumul} presents the cumulative cosmological mass density of neutral gas in DLAs 
as a function of the maximum \HI\ column density from data in our study and from the
different fits to the frequency distribution. 
It is apparent that $\omegagdla$ converges by $\log N(\HI)=22$. 

\begin{figure}
\includegraphics[width=\hsize,bb=75 175 515 586]{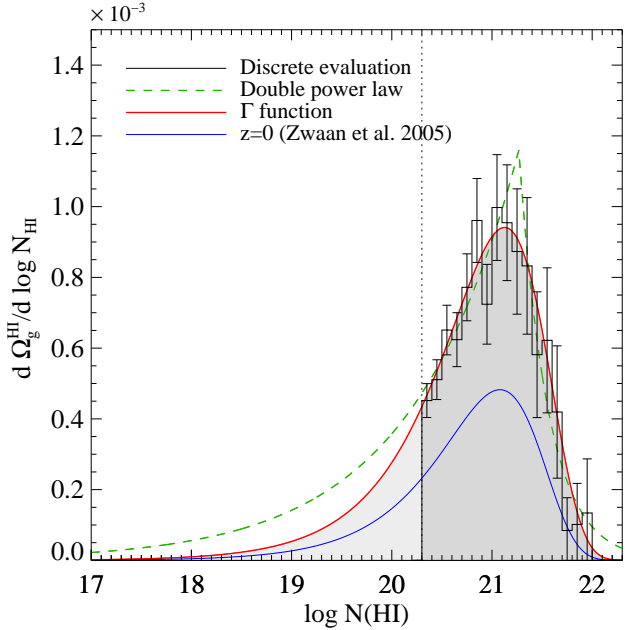}
\caption{
Cosmological mass density of neutral gas contained in systems of different column densities. 
The dashed-green and red curves correspond respectively to 
the Double power-law and $\Gamma$-function fits to $\fhix$ in SDSS DR7, while the blue solid curve 
represents the $\Gamma$-function fit to $\fhix(z=0)$ \citep{Zwaan05}.
\label{derivomegacumul}}
\end{figure}

A change of inflexion in the frequency distribution is apparent at $\log N(\HI)\sim 21$.
This is best seen on Fig.~\ref{derivomegacumul} which gives 
the slope of the above quantity as a function of $\log N(\HI)$. In other words, the area below the 
curve represents the contribution of the different intervals of $N(\HI)$ to the total \HI\ mass density. 
It is apparent that systems with very large (resp. low) column densities contribute little 
to the census of neutral gas because of their paucity (resp. low column density). 
The most important contribution comes from DLAs with $\log N(\HI)\sim21.2$.
A simple extrapolation of the $\Gamma$-function fit to column densities smaller than
$\log N(\HI)=20.3$ shows that sub-DLAs, with $19\la\log N(\HI)<20.3$, contribute about 20\% 
of the mass density of neutral hydrogen at $z\ga2.2$. Extrapolating the double 
power-law fit gives a sub-DLA contribution to $\Omega^{\HI}_{\rm g}$ of about 30\%.

Although both the power law and the Gamma function are good fits to $\fhix$, 
it can be seen on Figs.~\ref{fhix} and \ref{derivomegacumul} that the slope of 
$\fhix$ could be smaller at the low end ($\log N(\HI)\sim 20.3$) \citep{Peroux05, 
Guimaraes09}. The gamma function could best reproduce this regime. 
Furthermore, the double power-law produces a spike seen at $\log N(\HI)=21.3$, that 
is not present in the data. This is due to the arbitrary and somewhat unphysical discontinuous 
change in the slope of the double power-law fit. 

It is interesting to note here that the contribution of the different column densities 
to $\Omega^{\HI}_{\rm g}$ at high redshift
is very similar to what is observed in the local Universe \citep{Zwaan05}. 
This indicates that the \HI\ surface density profile of the neutral phase at high-$z$ is not 
significantly different from that observed at $z=0$.

\subsection{Evolution with cosmic time}

In this Section, we study the evolution over cosmic time of the cosmological mass
density of neutral gas, $\omegagdla$. This evolution is the result of several important 
processes involved in galaxy formation
including the consumption of gas during star formation activity but also
the consequences of energy releases during  
the hierarchical building up of systems from smaller blocks \citep[e.g.][]{Ledoux98, Haehnelt98},
or the ejection of gas from the central parts of massive halos
into the intergalactic medium through galactic winds \citep{Fall93}.

\citet{Storrie-Lombardi00} and \citet{Peroux03} claimed an increase of $\omegagdla$ when $z$ decreases
from $z\sim3$ to $z\sim2$. This is due to the lack in their sample of high column density DLAs at high redshift. 
From their SDSS DLA search, \citet{Prochaska05} observe on the contrary a significant decrease 
of $\omegagdla$ between $z\sim4$ and $z\sim2$. They interpret this as the result of neutral gas 
consumption by star formation activity and/or the ejection of gas into the intergalactic medium. The value they 
derive for $\omegagdla$ at $z=2.2$ is almost equal to that at $z=0$, indicating very little or no evolution 
of the cosmological mass density of neutral gas over the past ten billion years \citepalias[see][]{Prochaska09s}. 
We argue here that the differences between these results are artificial and can be reconciled 
within errors at least up to $z\sim3.2$. 

\begin{figure}
\includegraphics[width=\hsize,bb=75 175 515 586]{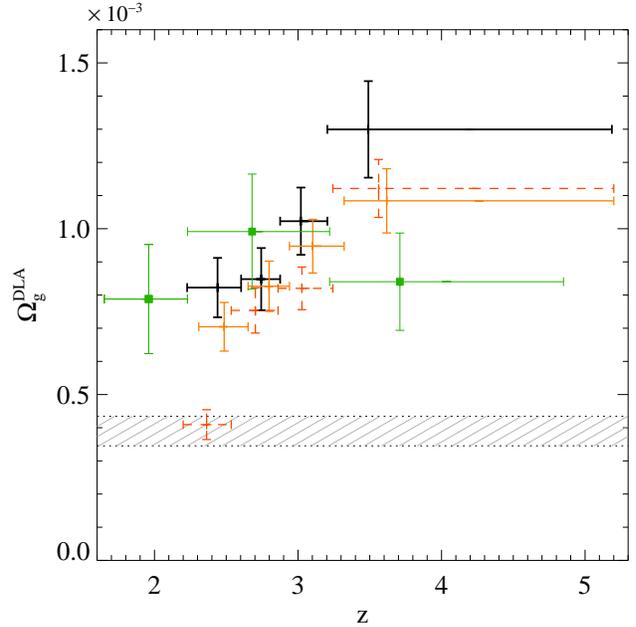}
\caption{The cosmological mass density of neutral gas in DLAs, $\omegagdla$, as a function of 
redshift for $z>1.65$. Vertical error bars represent the 1\,$\sigma$ uncertainties on the
measurements while horizontal error bars represent the redshift bins under consideration. 
Black error bars are our measurements at $z>2.2$. Orange error bars 
are derived from the \citetalias{Prochaska09s} sample when applying a $\delta {\rm v}=10\,000$~
\kms~cutoff to correct for the edge-effect (see Section~\ref{sbias} and Eq.~\ref{eqcorr}). 
The uncorrected values derived from the same sample are shown in dashed orange error bars. 
The green squares are derived from the sample of \citet{Peroux03} using a different binning 
compared to that adopted by these authors. Finally, the hashed region represent the 1\,$\sigma$ 
range on $\omegagdla$ at $z=0$ from \citet{Zwaan05}. We note also that the amount of baryons in stars 
at $z=0$ is $\Omega_{\star}=(2.5\pm1.3)\times10^{-3}$ \citep{Cole01}. \label{omegaa}}
\end{figure}

On the one hand, the redshift pathlength probed along SDSS lines of sight should be restricted so 
that the edge bias described in Section~\ref{bias} is avoided.
We applied to the \citetalias{Prochaska09s} sample the same velocity cut as 
for our sample and find, as expected, that both samples yield similar results
(see Fig.~\ref{omegaa}).
The differences between the measurements
from the two SDSS samples are within statistical error bars and are mainly 
due to slightly higher completeness and larger number statistics of our sample. 
The immediate implication of the edge-effect correction is that it cannot
be claimed that $\omegagdla$ does not evolve from $z=2.2$ to $z=0$.

\begin{figure}
\includegraphics[width=\hsize,bb=75 175 515 586]{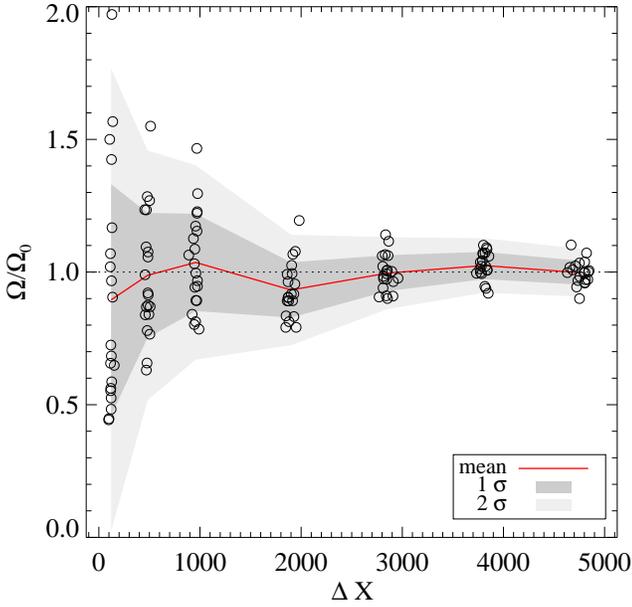}
\caption{Statistical uncertainty on the determination of $\omegagdla$ as a function of the 
sample size. Each circle represents the ratio between the measurement of $\omegagdla$ for a randomly selected 
sample of total absorption pathlength $\Delta X$ ($\Omega$) and that obtained from 
the full SDSS sample, $\Omega_{\rm o}$. 
Note that $\Delta X$~=~1\,540, 8\,475 and 11\,099 in,
respectively, \citet{Peroux03}, \citet[][after correcting for the edge-bias]{Prochaska09s} and the 
present work. \label{testssize}}
\end{figure}

On the other hand, we also note that a decrease of $\omegagdla$ with decreasing 
redshift cannot be excluded from the sample of \citet{Peroux03}. Indeed, a different binning of 
their data shows that they are actually consistent with
a decrease of $\omegagdla$ over the redshift range $z\sim$~3.2~$-$~2.2 (see Fig.~\ref{omegaa}).
In addition, while the sample from \citeauthor{Peroux03} is large enough to 
obtain a reasonable measurement of $\omegagdla$ at high redshift, it is probably too small to infer 
strong conclusions on its evolution. Fig.~\ref{testssize} illustrates the effect of the sample size 
on the determination of $\omegagdla$. We construct randomly selected SDSS QSO sub-samples of increasing 
total path length, $\Delta X$, and calculate the ratio $\Omega/\Omega_{\rm o}$ of 
$\omegagdla$ values from the subsample and the whole SDSS sample.
Note that the survey used in the present paper has $\Delta X$~=~11\,099 whereas \citeauthor{Prochaska09s}'s 
survey has $\Delta X$~=~8\,475 (applying the new definition of $\zmin$) and \citeauthor{Peroux03}'s survey has 
$\Delta X$~=~1\,540.

\begin{figure}
\includegraphics[width=\hsize,bb=75 175 515 605]{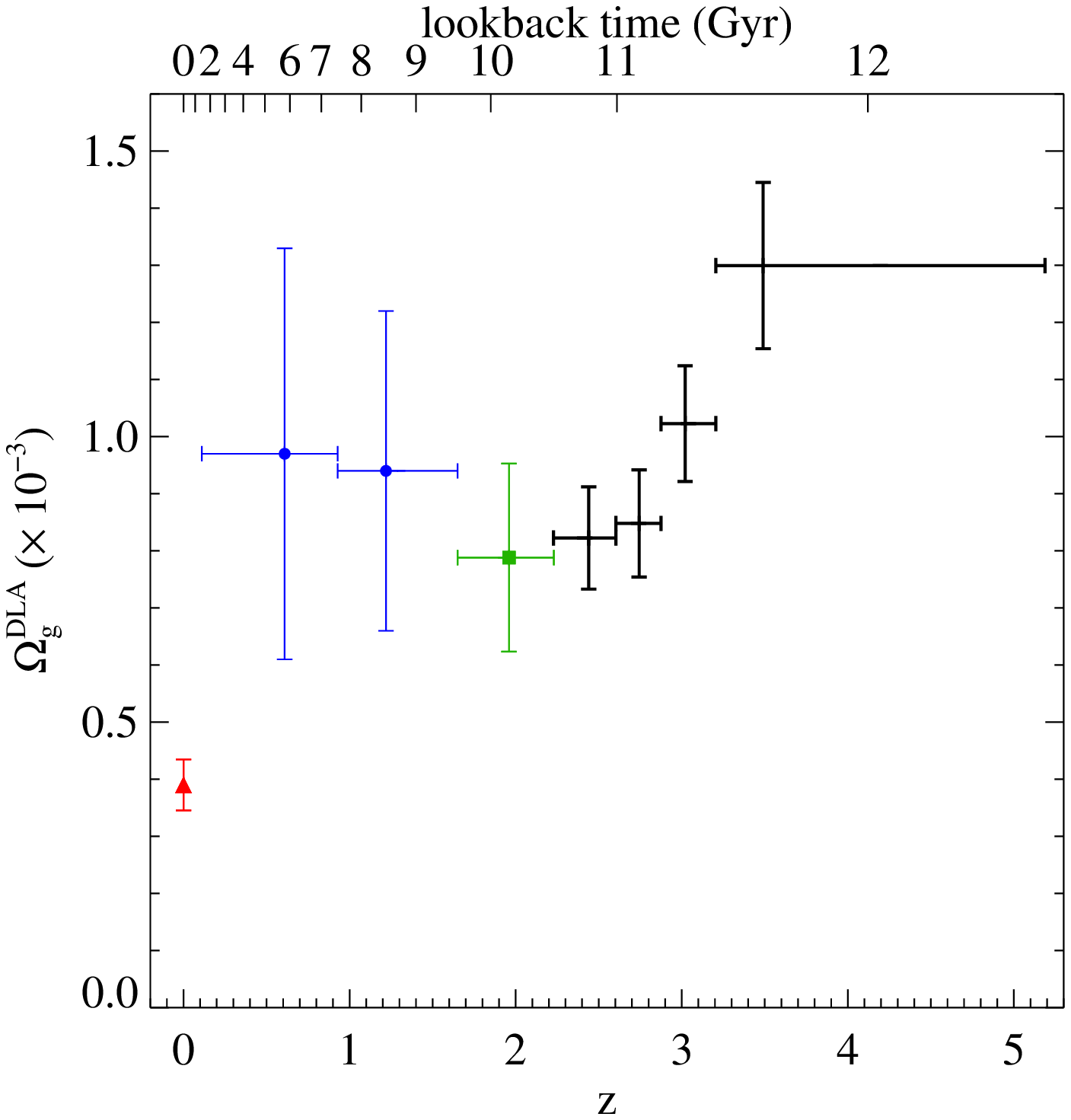}
\caption{\label{omega} Cosmological mass density of neutral gas in DLAs, $\omegagdla$, as a function of 
redshift. The red triangle at $z=0$ is the value from 21-cm maps by \citet{Zwaan05}. The blue filled 
circles at $z\sim1$ are the measurements of $\omegagdla$ from \citet{Rao06}. 
The green square at $z\sim 2$ is derived from the sample of \citet{Peroux03}. Measurements at $z>2.2$ 
are from the present work base on SDSS DR7.}
\end{figure}

All this reconciles the SDSS measurements with the results from \citeauthor{Peroux03} 
(see Fig.~\ref{omega}). It can be seen also that the points at $z\sim 2$ are consistent with
those at lower redshift given their large error bars \citep{Rao06}. Note also that the value 
we derive at $z\sim2.4$ is consistent with that obtained from the Hamburg-ESO survey 
\citep[$\omegagdla\sim1$, A. Smette, {\sl private communication}; ][]{Smette05}. 

At redshifts above $z=3.2$, there is still a discrepancy between the results from SDSS and that 
from previous surveys (see Fig.~\ref{omegaa}). As pointed out by \citet{Prochaska05}, the size of 
the samples prior to SDSS were insufficient to detect the convergence of $\omegagdla$ at large 
$N(\HI)$. Therefore, the inclusion of a single large column density system in these samples could change 
the results on $\omegagdla$ significantly. In turn, the low spectral resolution of SDSS combined 
with a dense \lya\ forest could 
lead to slightly overestimate the column densities of high-$z$ DLAs. 

The measurement of the cosmological mass density of neutral gas at intermediate and low redshifts is a 
difficult task. The little incidence of DLAs and the need for observations from space have lead to samples 
of limited sizes. \citet{Rao00} and \citet{Rao06} used a novel technique to measure $\omegagdla$ 
at $0.1<z<1.6$. They searched for the \lya\ absorption associated to \MgII\ systems, which statistics is very large 
at those redshifts. The corresponding values of $\omegagdla$ are high and therefore have been extensively discussed in the literature.
In particular, \citetalias{Prochaska09s} concluded that the values obtained at $z\sim1$ by \citet{Rao06} 
are difficult to reconcile with the value they obtained at $z\sim2.2$ and suffer 
from a statistical fluke or an observational bias. After correcting for the edge bias we discussed
earlier it can be seen that the SDSS results are no more incompatible with the \citet{Rao06} results.
However, it is still possible that the high values of $\omegagdla$ from \citeauthor{Rao06} are overestimated 
due to systematics associated with the selection of low and intermediate redshift DLAs directly 
from strong \MgII\ absorption \citep[see][]{Peroux04, Dessauges-Zavadsky09}. 
Only a large blind survey for DLAs at $z\sim1$ could solve this issue.

Note also that $\omegagdla(z\sim2)$, 
obtained from the sample of \citet{Peroux03}, is almost equal to $\omegagdla(z\sim2.2)$, obtained here from SDSS-DR7. 
This could indicate a flattening of $\omegagdla(z)$ at $z \sim 2$. Large intermediate-redshift 
optical and radio surveys are therefore still required to constrain the evolution of $\omegagdla$ from $z=2$ to $z=0$ 
\citep[see discussion in ][]{Gupta09}.

\begin{table}
\caption{The cosmological mass density of neutral gas in DLAs from this work (SDSS-DR7)\label{omegat}}
\begin{center}
\begin{tabular}{c c c l}
\hline
\hline
$z$        & $\avg{z}^a$  & $\Delta X$ & $\omegagdla$ ($\times 10^{-3}$) \\
\hline     
2.23-2.60  & 2.44  &  2774      &   0.82$\pm$0.09           \\
2.60-2.88  & 2.74  &  2774      &   0.85$\pm$0.09           \\
2.88-3.20  & 3.02  &  2774      &   1.03$\pm$0.10           \\
3.20-5.19  & 3.49  &  2774      &   1.29$\pm$0.15           \\
\hline
\end{tabular}
\end{center}
$^a$ median redshift corresponding to half the total pathlength $\Delta X$ in the redshift bin.
\end{table}

\section{Conclusion \label{concl}}

We have demonstrated the feasibility and the robustness of a fully automatic search of SDSS-DR7 for 
DLA systems based on the identification of DLA profiles by correlation analysis. 
This led to the identification of about one thousand DLAs, representing the largest DLA 
database to date. 
We tested the accuracy of the $N(\HI)$ measurements and quantified the high level of completeness and reliability 
of the detections. 

In agreement with previous studies \citep{Peroux03,Prochaska05,Prochaska09s}, we show that a single power-law 
is a poor description of the $N(\HI)$-frequency distribution at $\log N(\HI)\ge20.3$. 
A double power-law or a $\Gamma$ function give better fits. The finding of one $\log N(\HI)=22$ DLA, 
confirmed by UVES high spectral resolution observations, shows that the slope of $\fhix$ at $\log N(\HI)>21.5$ is $-3.5$.

The convergence of $\omegagdla$ for large $N(\HI)$ indicates that the cosmological mass density of 
neutral gas at $z\sim2.2-5$ is dominated by bona-fide damped Lyman-$\alpha$ 
systems. The relative contribution of DLAs reaches its maximum around $\log N(\HI)=21$, similar to what is 
observed in the local Universe. The paucity of very high-column density DLAs implies that they contribute 
for only a small fraction to the cosmological mass density of neutral gas. On the other hand, an extrapolation of $\fhix$ at 
$\log N(\HI)<20.3$ suggests that sub-DLA systems contribute to about one fifth of the neutral hydrogen at 
high redshift, in agreement with the results of \citet{Peroux05}.

We identified an important observational bias due to an edge effect 
and proposed a method to avoid it. Such a bias could also partly explain the higher 
values of $\omegagdla$ found by \citet{Prochaska05} when selecting only bright quasars, as 
the bias discussed here preferentially affects faint quasars with lower signal-to-noise ratios. 
Indeed, when not correcting for the bias, we find 10\% higher $\omegagdla$ from a bright QSO 
sub-sample ($i<19.5$) compared to a faint QSO sub-sample ($i\ge 19.5$) while this difference is only 
5\% when the bias is avoided. 

We derive the evolution with time of the cosmological mass density of neutral gas in Fig.~\ref{omega} 
and summarise our measurements in Table~\ref{omegat}.
We observe a decrease with time of the cosmological mass density of neutral gas between $z\sim3.2$ and $z\sim2.2$, 
confirming the results from \citet[also \citealt{Prochaska09s}]{Prochaska05}. However, we argue that the value at 
$z\sim2.2$ is significantly higher (by up to a factor of two) than the value at $z=0$, indicating that 
$\omegagdla$ keeps evolving at $z<2.2$. Interestingly, models of the evolution of the reservoir of neutral gas \citep{Hopkins08} 
also predict a value of $\omegagdla$ at $z\sim 2.2$ higher than that at $z=0$. 
The small number statistics at high redshift and the insufficient spectral resolution of SDSS spectra do 
not allow for a strong conclusion on the neutral gas content of the Universe at $z\sim4-5$.  
Further surveys are therefore required at $z>4$ \citep[see e.g.][]{Guimaraes09}. 

We measure $\omegagdla(z\sim3)\approx10^{-3}$. This implies that neutral gas accounts for 
only 2\% of the baryons at high redshift, according to the latest cosmological parameters from WMAP 
\citep{Komatsu09}. This implies that most of the baryons are in the form of ionised gas in the intergalactic 
medium \citep[e.g.][]{Petitjean93}. 

The amount of baryons locked up into stars at $z=0$ \citep[$\Omega_\star=(2.5\pm1.3)\times10^{-3}$;][]{Cole01} is 
about twice the amount of neutral gas contained in high redshift DLAs. This implies that the DLA phase must be 
replenished in gas before the present epoch \citepalias[see also][]{Prochaska09s} at a rate similar to 
that of its consumption \citep{Hopkins08}. 
This is also required to explain the properties of $z=2-3$ Lyman-break galaxies \citep{Erb08}.
The replenishment of \HI\ gas could take place through the accretion of matter from the intergalactic 
medium and/or recombination of ionised gas in the walls of supershells. 
Several observational evidences of cold gas accretion at high redshift have been published recently 
\citep[e.g.][]{Nilsson06,Dijkstra06,Noterdaeme08hd}. On the other hand, supershells provide a natural 
explanation to the proportionality between star formation and replenishment rate \citep{Hopkins08}.
The results presented here provide strong constraints for numerical modelling of hierarchical evolution 
of galaxies \citep[e.g.][]{Pontzen08}. Note that galactic winds are likely to play an 
important role in the evolution of the cosmological mass density of neutral gas \citep{Tescari09}.

Finally, it has long been discussed whether the optically selected quasar samples are affected by extinction 
due to the presence of dust on the line of sight \citep[][]{Boisse98, Ellison01, Smette05}. We recently 
presented direct evidence that lines of sight towards colour-selected quasars are biased against the detection 
of diffuse molecular clouds \citep{Noterdaeme09co}. \citet{Pontzen09} estimated that 
dust-biasing could lead to underestimate the metal budget by about 50\%. 
Although it has been claimed that the global census of neutral gas should be little affected by dust-biasing 
\citep{Ellison08,Trenti06}, it will be interesting to revisit this issue.

\begin{acknowledgements}
We thank the anonymous referee for useful suggestions and comments.
We thank Jason Prochaska for proving us with his results on SDSS-DR5 prior to publication. 
We are also grateful to Alain Smette for communicating us his measurements on the 
Hamburg-ESO survey.
PN acknowledges support from the french Ministry of Foreign and European Affairs. 
We acknowledge the tremendous effort put forth by the Sloan Digital Sky Survey team
to produce and release the SDSS survey. Funding for the SDSS and SDSS-II has been 
provided by the Alfred P. Sloan Foundation, the Participating Institutions, the 
National Science Foundation, the U.S. Department of Energy, the National Aeronautics 
and Space Administration, the Japanese Monbukagakusho, the Max Planck Society, and 
the Higher Education Funding Council for England. The SDSS Web Site is http://www.sdss.org/.
The SDSS is managed by the Astrophysical Research Consortium for the Participating Institutions. 
The Participating Institutions are the American Museum of Natural History, Astrophysical 
Institute Potsdam, University of Basel, University of Cambridge, Case Western Reserve 
University, University of Chicago, Drexel University, Fermilab, the Institute for Advanced 
Study, the Japan Participation Group, Johns Hopkins University, the Joint Institute for 
Nuclear Astrophysics, the Kavli Institute for Particle Astrophysics and Cosmology, the 
Korean Scientist Group, the Chinese Academy of Sciences (LAMOST), Los Alamos National 
Laboratory, the Max-Planck-Institute for Astronomy (MPIA), the Max-Planck-Institute for 
Astrophysics (MPA), New Mexico State University, Ohio State University, University of 
Pittsburgh, University of Portsmouth, Princeton University, the United States Naval 
Observatory, and the University of Washington.
\end{acknowledgements}

\bibliographystyle{aa}
\bibliography{../../mybib}

\end{document}